\documentclass[letterpaper,american,english,reprint, aps]{revtex4-1}
\usepackage[T1]{fontenc}
\usepackage[latin9]{inputenc}
\usepackage{color}
\usepackage{amsmath}
\usepackage{amssymb}
\usepackage{graphicx}
\usepackage{filecontents}
\usepackage{natbib}

\makeatletter

\pdfpageheight\paperheight
\pdfpagewidth\paperwidth

\definecolor{blue}{cmyk}{0, 0, 0, 1}

\makeatother

\usepackage{babel}

\begin{document}

\preprint{APS/123-QED}

\title{Quantum optical realization of arbitrary \\
linear transformations allowing for loss and gain}

\author{N. Tischler}
\email{n.tischler@griffith.edu.au}

\affiliation{Centre for Quantum Dynamics, Griffith University, Brisbane 4111,
Australia}

\affiliation{Department of Physics and Astronomy, Macquarie University, Sydney
2109, Australia}

\affiliation{Vienna Center for Quantum Science and Technology, Faculty of Physics,
University of Vienna, Boltzmanngasse 5, A-1090 Vienna, Austria}

\author{C. Rockstuhl}
\email{carsten.rockstuhl@kit.edu}

\affiliation{Institute of Theoretical Solid State Physics, Karlsruhe Institute
of Technology, 76131 Karlsruhe, Germany}

\affiliation{Institute of Nanotechnology, Karlsruhe Institute of Technology, 76021
Karlsruhe, Germany }

\author{K. S\l owik}
\email{karolina@fizyka.umk.pl}

\affiliation{Institute of Physics, Faculty of Physics, Astronomy and Informatics,
Nicolaus Copernicus University, Grudziadzka 5, 87-100 Torun, Poland}

\date{\today}
\begin{abstract}
Unitary transformations are routinely modeled and implemented in the
field of quantum optics. In contrast, nonunitary transformations that
can involve loss and gain require a different approach. In this theory
work, we present a universal method to deal with nonunitary networks.
An input to the method is an arbitrary linear transformation matrix
of optical modes that does not need to adhere to bosonic commutation
relations. The method constructs a transformation that includes the
network of interest and accounts for full quantum optical effects
related to loss and gain. Furthermore, through a decomposition in
terms of simple building blocks it provides a step-by-step implementation
recipe, in a manner similar to the decomposition by Reck \textit{et
al.} \citep{Reck1994} but applicable to nonunitary transformations.
Applications of the method include the implementation of positive-operator-valued
measures and the design of probabilistic optical quantum information
protocols.
\end{abstract}
\maketitle

\section{\label{sec:level1}Introduction\protect \\
 }

Transformations between sets of orthogonal input and output modes
are ubiquitous in optics and quantum information technology. In particular,
linear transformations between the amplitudes of the input and output
modes are used to perform a variety of tasks, e.g. to operate single
qubit gates or to model the action of physical elements such as beam
splitters \footnote{By the terms `linear transformation' and `linear network' we refer
to transformations for which the expectation values of the fields
are related by a linear transformation between the input and output
modes and the annihilation operators of the output modes have the
same linear dependence on the input annihilation operators. }. Mathematically, a linear transformation can be expressed as a transformation
matrix $T$ relating the mean fields of the $m$ optical input modes
$1\mathrm{in}...m\mathrm{in}$ with those of the $n$ optical output
modes $1\mathrm{out}...n\mathrm{out}$:
\begin{equation}
\left(\begin{array}{c}
\begin{array}{c}
\left\langle \hat{a}_{1\mathrm{out}}\right\rangle \\
\vdots
\end{array}\\
\left\langle \hat{a}_{n\mathrm{out}}\right\rangle 
\end{array}\right)=T\left(\begin{array}{c}
\begin{array}{c}
\left\langle \hat{a}_{1\mathrm{in}}\right\rangle \\
\vdots
\end{array}\\
\left\langle \hat{a}_{m\mathrm{in}}\right\rangle 
\end{array}\right).\label{eq:Tmeanfields}
\end{equation}

Among such transformations, unitary optical networks, for which $T$
is a unitary matrix that also relates the annihilation operators themselves
and not only their expectation values, are routinely used in optical
quantum information processing. Unitary networks conserve the number
of photons and their implementation in terms of basic building blocks,
namely phase shifters acting on individual modes and beam splitters
mixing two modes at a time, is well understood \citep{Reck1994,Clements2016}.
However, as unitarity imposes restrictions on the transformation matrix,
unitary networks can be considered as a special case of linear networks. 

Relaxing the restrictions unlocks fascinating opportunities for new
transformations, including the options of loss and gain \citep{Barnett1998,Knoell1999,Jeffers2000,Scheel2000,Lee2012,Miller2013,Miller2012,Gupta2014,Roger2015,Roger2016,Uppu2016}.
One noteworthy class of such networks consists of asymmetric nonunitary
beam splitters, which can allow highly tunable quantum interference
\citep{Uppu2016}. Among the symmetric beam splitters, an example
of a nonunitary beam splitter that has attracted particular interest
is the $2\times2$ transformation given by the matrix $T=\frac{1}{2}\left(\begin{array}{cc}
1 & -1\\
-1 & 1
\end{array}\right)$ \citep{Barnett1998,Jeffers2000,Lee2012,Roger2015,Roger2016}. A device
with this action can be thought of as a lossy beam splitter. It exhibits
a striking, apparently nonlinear, behavior when one photon is incident
on each input: either both photons are or neither of them is lost. 

Even though the initial interest in devices such as this one was primarily
theoretical, the technical capabilities in the design and fabrication
of novel and nanostructured materials are now making elements with
such properties possible \citep{Roger2015,Roger2016,DiMartino2014,Cai2014,Fujii2014,Fakonas2015}.
Nonunitary transformation matrices also prove useful in modeling the
inevitable imperfections of real optical elements that show a wavelength
dependent behavior \citep{Barnett1998}. A further reason for stepping
outside the framework of unitary networks is that transformations
may have an unequal number of input and output modes of interest,
a clear indicator of nonunitarity. Two particularly simple examples
are Y-junctions in integrated optics and absorptive polarizers, which
feature two orthogonal input modes but only one output mode. 

For a quantum optical description of such transformations, the relationship
of Eq. (\ref{eq:Tmeanfields}) does not suffice. Additionally, a relationship
between annihilation and creation operators is required. It would
be tempting to simply drop the expectation values in Eq. (\ref{eq:Tmeanfields}),
but the modes associated with nonunitary networks would generally
not fulfill the required bosonic commutation relations. Hence, we
will from now on drop the expectation values and take $T$ to be a
transformation between the annihilation operators of interest with
the understanding that it is an incomplete transformation: ancilla
modes need to be introduced in the mathematical description to faithfully
reproduce or predict the full quantum optical transformation. Although
this is straightforward for the simple examples of Y-junctions and
polarizers, a systematic method to deal with larger scale problems
would be desirable. 

In this paper we investigate whether such a strategy is possible for
all linear transformation matrices, how many ancilla modes are needed
for any given case, and how a full enlarged quantum optical network
can be mathematically represented and physically realized. 

Related problems have been previously studied in a number of works.
In Refs. \citealp{Miller2013} {\color{blue} and \citealp{Miller2012}} Miller shows how to construct universal linear transformation machines in a classical optics picture where the mean fields are of interest, so that a modulation of field amplitudes is possible without the need to take into account quantum optical effects. The Bloch-Messiah reduction also shows how a decomposition
into basic building blocks can be found and it does include a rigorous
quantum optical description. However, it already starts with the complete
transformation matrix respecting bosonic commutation relations (a
linear unitary Bogoliubov transformation), rather than a partial network
\citep{Loock2011}. Allowing nonunitary partial networks as an input,
{\color{blue} He \emph{et al.} and} Kn\"{o}ll and coworkers present {\color{blue} techniques} to find corresponding enlarged
{\color{blue} transformations}, but {\color{blue} they} do not allow for transformations that include
both loss and gain  \citep{He2007,Knoell1999,Scheel2000}. 

In this article we put forward a systematic method for dealing with
linear transformation matrices of any size, allowing for the option
of loss and gain. The method combines a singular value decomposition
of the partial network and the single mode treatment presented in
Ref. \citealp{Leonhardt2003} to provide full information about the
transformation, so that the quantum optical output state can be calculated
for any input state. In addition, as a generalization of the seminal
decomposition in Ref. \citealp{Reck1994} or the more recent variant
of Ref. \citealp{Clements2016} to nonunitary networks, our method
shows how to realize transformations in terms of the basic building
blocks of phase shifters, beam splitters, and parametric amplifiers. 

We discuss possible applications of nonunitary networks, which include
the implementation of positive-operator-valued measures (POVMs) and
probabilistic optical quantum information protocols. The physical
realization of small circuits could be achieved with bulk optics,
whereas integrated optics would be naturally suited as a platform
for larger scale networks. In the appendix we demonstrate the method
on several examples, including the lossy beam splitter with apparent
nonlinear action described earlier. The lossy beam splitter example
illustrates how devices made of exotic materials can be replaced by
standard optical circuits. 

\section{Results\label{sec:Results}}

\begin{figure}
\includegraphics{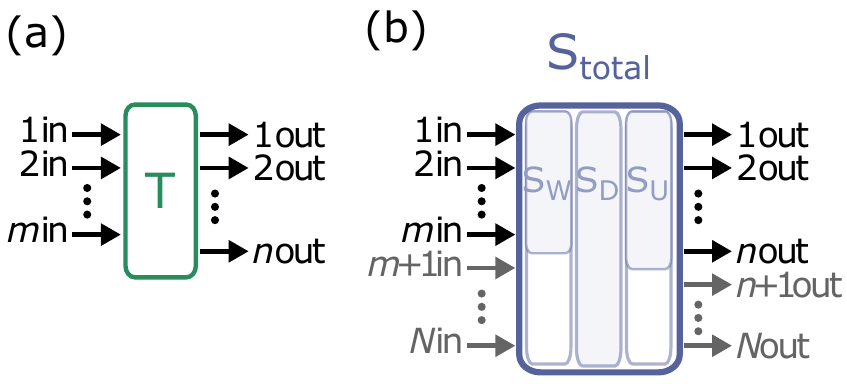}

\caption{The concept of mode transformations. (a) The linear network $T$ specifies
a mapping from $m$ input modes to $n$ output modes and may be characterized
by a nonunitary matrix. (b) The full network $S_{\mathrm{total}}$
includes the nominal modes of $T$, as well as ancilla modes, which
account for any losses and gain in $T$. The transformation $S_{\mathrm{total}}$
consists of three main components, of which only the second involves
a coupling between the nominal modes and ancilla modes.\label{fig:ModeTransformations}}
\end{figure}

We begin by outlining the basic structure of the method, {\color{blue} illustrated in Fig. \ref{fig:ModeTransformations}.} Starting
with the partial network $T$, a singular value decomposition is performed,
which yields three main components, {\color{blue} $U$, $D$, and $W$. The singular value decomposition is particularly useful as each main component is well suited to be further decomposed into a sequence of
operations in the form of simple building blocks. Each of these building blocks corresponds to a physical operation and has a
known complete quantum optical description.}

{\color{blue} Importantly, since $U$ and $W$ are unitary, they can physically be implemented with phase shifters and beam splitters using the techniques of Refs. \citealp{Reck1994} or \citealp{Clements2016}.} These two main components only involve the nominal modes and can be understood
as an initial conversion from the input modes to another basis, the modulation
basis, and a final conversion from the modulation basis to the output
modes. The modulation takes place in {\color{blue}$D$}, the second main component, and
includes interactions with ancilla modes. {\color{blue} Specifically, each operation here corresponds to a singular value, and each singular value different from one results in the interaction of a nominal mode with a vacuum ancilla, either through a beam splitter or a parametric amplifier.}

Combining all of the individual
operations provides the quantum optical description of the overall transformation,
which we denote by $S_{\mathrm{total}}$.

\subsection{Preliminaries}

As a basis for the detailed description of the method in Section \ref{subsec:Method},
it is useful to first establish some terminology and a single-mode
framework after Ref. \citealp{Leonhardt2003}, i.e. the case with
a single nominal input mode and a single nominal output mode. {\color{blue} In the general multi-mode treatment put forward in the present article, we will make extensive use of these basic single-mode tools.}

\subsubsection{Quasiunitarity}

A $2N\times2N$-dimensional matrix $S$ is quasiunitary if 

\begin{equation}
SGS^{\dagger}=G,\label{eq:quasiunitarity}
\end{equation}
where $G$ is defined as the $2N\times2N$ diagonal matrix with the
first $N$ diagonal elements equal to $1$ and the last $N$ diagonal
elements equal to $-1$ \citep{Williamson1937,Leonhardt2003}.

\subsubsection{Properties of partial and full transformations}

\begin{figure}
\includegraphics{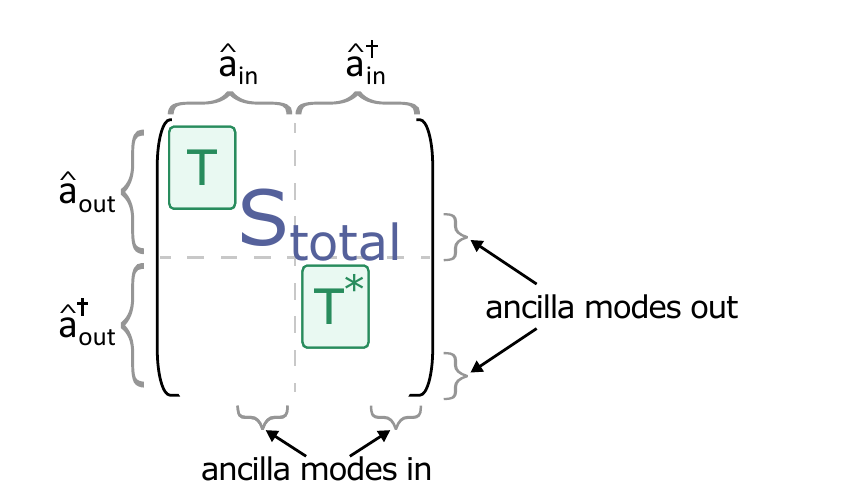}

\caption{The structure of the matrix $S_{\mathrm{total}}$. The matrix elements
represent the coupling between the respective input modes (columns)
and output modes (rows). Viewed as a $2\times2$ block matrix, non-zero
elements in the off-diagonal blocks are responsible for active elements
in the implementation. The transformation of interest, $T$, is contained
in the upper left block. \label{fig:Sstructure}}
\end{figure}

The input into the method is the partial network $T$, a complex matrix
of any size without any conditions on its elements. We call $T$ a
partial network because in general, $T$ on its own is not enough
to predict the quantum optical output for an arbitrary input. For
instance, the noise due to vacuum fluctuations in ancilla modes is
neglected, and this noise impacts quantum properties of light such
as the degree of squeezing. One of the aims of the method is to construct
another network, $S_{\mathrm{total}}$, which contains $T$ as its
upper left block and includes the ancilla modes so that it can be
used as a quantum optical model of the transformation $T$ (Fig. \ref{fig:Sstructure}).
The matrix $S_{\mathrm{total}}$ relates the input and output creation
and annihilation operators in the following way:
\begin{equation}
\left(\begin{array}{c}
\begin{array}{c}
\hat{a}_{1\mathrm{out}}\\
\vdots
\end{array}\\
\hat{a}_{N\mathrm{out}}\\
\hat{a}_{1\mathrm{out}}^{\dagger}\\
\vdots\\
\hat{a}_{N\mathrm{out}}^{\dagger}
\end{array}\right)=S_{\mathrm{total}}\left(\begin{array}{c}
\begin{array}{c}
\hat{a}_{1\mathrm{in}}\\
\vdots
\end{array}\\
\hat{a}_{N\mathrm{in}}\\
\hat{a}_{1\mathrm{in}}^{\dagger}\\
\vdots\\
\hat{a}_{N\mathrm{in}}^{\dagger}
\end{array}\right).
\end{equation}
It is $2N\times2N$-dimensional, where in general $N\geq\max\left(m,n\right)$
due to the possible inclusion of ancilla modes. A requirement on $S_{\mathrm{total}}$
is that it must fulfill the quasiunitarity equation (\ref{eq:quasiunitarity})
so that its modes are bosonic, i.e. the creation and annihilation
operators fulfill the standard bosonic commutation relations $\left[\hat{a}_{i},\hat{a}_{j}\right]=0$,
$\left[\hat{a}_{i},\hat{a}_{j}^{\dagger}\right]=\delta_{ij}$. The
reason that creation operators are included in the description is
that active elements associated with gain lead to a coupling of creation
and annihilation operators. In fact, whether the transformation contains
only passive elements or includes active elements can be recognized
based on the off-diagonal blocks of $S_{\mathrm{total}}$ when viewed
as a $2\times2$ block matrix: a passive transformation has zeros
for these blocks.

\subsubsection{Single-mode loss }

A single lossy channel characterized by $T=\sigma$ where $\sigma\in\mathbb{R},$
$0\le\sigma<1$, {\color{blue} can be implemented using a lossless beam splitter 
with an ancilla mode $\hat{a}_{2}$ initialized in its vacuum state. 
The transformation of the modes is then generated by a beam splitter Hamiltonian
\mbox{$\hat{H} = i\phi \left(\hat{a}_1^\dagger \hat{a}_2 - \hat{a}_2^\dagger \hat{a}_1 \right)$}, with $\cos \phi = \sigma$ and \mbox{$\sin\phi = \sqrt{1-\sigma^2}$} representing the transmission and reflection amplitudes of the beam splitter, respectively. 
The connection between this Hamiltonian and the corresponding transformation matrix }
\begin{equation}
S=\left(\begin{array}{cccc}
\sigma & \sqrt{1-\sigma^{2}} & 0 & 0\\
-\sqrt{1-\sigma^{2}} & \sigma & 0 & 0\\
0 & 0 & \sigma & \sqrt{1-\sigma^{2}}\\
0 & 0 & -\sqrt{1-\sigma^{2}} & \sigma
\end{array}\right)\label{eq:S_BS}
\end{equation}
{\color{blue} such that }
\begin{equation}
\left(\begin{array}{c}
\hat{a}_{1\mathrm{out}}\\
\hat{a}_{2\mathrm{out}}\\
\hat{a}_{1\mathrm{out}}^{\dagger}\\
\hat{a}_{2\mathrm{out}}^{\dagger}
\end{array}\right)=S\left(\begin{array}{c}
\hat{a}_{1\mathrm{in}}\\
\hat{a}_{2\mathrm{in}}\\
\hat{a}_{1\mathrm{in}}^{\dagger}\\
\hat{a}_{2\mathrm{in}}^{\dagger}
\end{array}\right),
\end{equation}
{\color{blue} is described in Ref.~\onlinecite{Leonhardt2003}, pp.~1215--1216. (see also \footnote{We use a slightly different definition for the matrix $H$ in this connection: $H=iG \mathrm{ln}S$.}) }


\subsubsection{Single-mode gain}

Similarly, for a single channel with gain given by $T=\sigma$ where
$\sigma\in\mathbb{R},$ $\sigma>1$, we introduce an ancilla mode
$\hat{a}_{2}$, {\color{blue} initially in vacuum.
Gain can be realized with a parametric amplifier with a Hamiltonian 
$\hat{H} = i\xi \left(\hat{a}^\dagger_1\hat{a}^\dagger_2 - \hat{a}_1\hat{a}_2 \right)$,
where $\cosh \xi = \sigma$. 
The corresponding enlarged transformation  }
\begin{equation}
S=\left(\begin{array}{cccc}
\sigma & 0 & 0 & \sqrt{\sigma^{2}-1}\\
0 & \sigma & \sqrt{\sigma^{2}-1} & 0\\
0 & \sqrt{\sigma^{2}-1} & \sigma & 0\\
\sqrt{\sigma^{2}-1} & 0 & 0 & \sigma
\end{array}\right)
\end{equation}
{\color{blue}can be constructed as described in Ref.~\onlinecite{Leonhardt2003}, p.~1217.
}

\subsubsection{Single-mode phase shift}

Complex transformations involve phase shifts $\mbox{T=\ensuremath{e^{i\varphi}}}$
($\varphi\in\mathbb{R}$) for which no ancilla mode is required.
{\color{blue} The Hamiltonian is $\hat{H} = -\varphi \hat{a}_1^\dagger\hat{a}_1$,
and the} transformation takes the simple form of $\mbox{\ensuremath{\left(\begin{array}{c}
 \hat{a}_{1\mathrm{out}}\\
 \hat{a}_{1\mathrm{out}}^{\dagger} 
\end{array}\right)}=S\ensuremath{\left(\begin{array}{c}
 \hat{a}_{1\mathrm{in}}\\
 \hat{a}_{1\mathrm{in}}^{\dagger} 
\end{array}\right)}}$ with
\begin{equation}
S=\left(\begin{array}{cc}
e^{i\varphi} & 0\\
0 & e^{-i\varphi}
\end{array}\right).
\end{equation}

\subsection{\label{subsec:Method}Method}

The method consists of the steps illustrated in Fig.\ \ref{fig3} and described below:

\begin{figure}[t]
\includegraphics{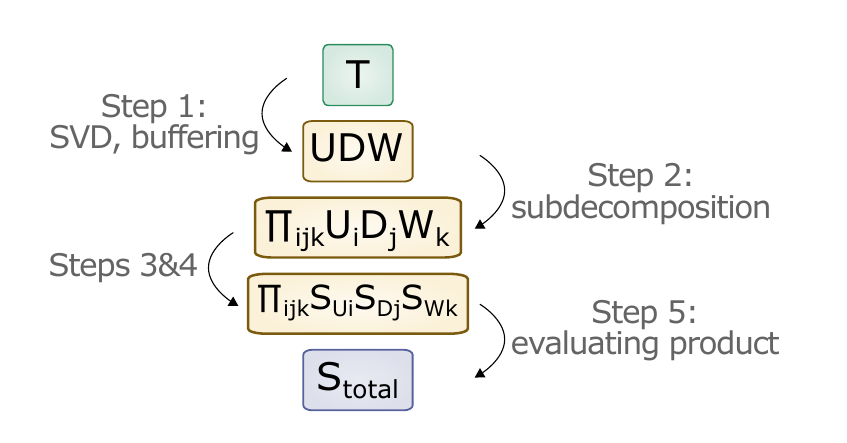}

\caption{Overview of the method. Starting with $T$, a singular value decomposition
followed by a padding with identity matrix elements to make all components
the same size, provides the main components $U$, $D,$ and $W$.
In Step $2$, a further decomposition of $U$ and $W$ into simple
blocks is found through the technique in Ref. \citealp{Reck1994}
or Ref. \citealp{Clements2016}, and $D$ is written as a product
of single-mode modulations. For each of these, a corresponding enlarged
transformation is determined in Steps $3$ and $4$. Finally, the
overall transformation $S_{\mathrm{total}}$ is obtained as the product
of all the individual transformations. \label{fig3}}
\end{figure}

\paragraph{Step 1, singular value decomposition of $T$: }

A singular value decomposition provides the main components 
\begin{equation}
T=UDW,\label{eq:svd}
\end{equation}
where $U$ and $W$ are unitary matrices and $D$ is a diagonal matrix
with non-negative real diagonal elements.

\paragraph{Step 1b, if $T$ is not square: }

The method can be applied to transformations of arbitrary dimensionality,
including those given by non-square matrices. Such transformations
apparently correspond to unequal numbers of input and output modes,
which is an incomplete description in quantum mechanics as it can
neglect necessary sources of quantum noise. For this reason, non-square
transformations definitely require either ancilla input modes or ancilla
output modes, so that the number of inputs matches the outputs. On
top of that, both square and non-square transformations may require
what we will refer to as full ancilla modes, which will be discussed
later on. 

A singular value decomposition of a non-square $n\times m$ matrix
provides a square $n\times n$ matrix $U$, a diagonal $n\times m$
matrix $D$, and another square $m\times m$ matrix $W$. The impact
of the missing input or output modes can be naturally taken into account
through augmentation of {\color{blue} the matrices $U$, $D$, and $W$} to
the $\max\left(m,n\right)\times\max\left(m,n\right)$ size, by padding
them with the corresponding elements of the identity matrix
as the last rows and columns where required. The following steps 2-5
should be applied to the augmented matrices, which we will still call
$U,$ $D,$ $W$ for simplicity. 

An example of an application of the method to a non-square matrix
is shown in Appendix \ref{subsec:Simple-POVM}.

\paragraph{Step 2, subdecomposition of all three matrices:}

We further decompose the two unitary matrices $U$ and $W$ by the
established methods of Ref. \citealp{Reck1994} or Ref. \citealp{Clements2016},
and thereby write the main components as the products $U=\prod_{i}U_{i}$,
and $W=\prod_{k}W_{k}$, respectively. {\color{blue} All of the matrices $U_{i}$ and $W_{k}$ correspond to simple physical operations of phase shifters and beam splitters.} The diagonal matrix $D$ can
be decomposed into a product of matrices $D=\prod_{j}D_{j}$, where
each $D_{j}$ is the identity matrix with element $\left(j,j\right)$
replaced by $D_{j,j}$. Overall, we obtain $T=\prod_{ijk}U_{i}D_{j}W_{k}$.

\paragraph{Step 3, determining the dimensionality of the enlarged system and
assigning modes: }

The dimensionality of the enlarged matrices is $2N\times2N$, with
$N$ given by $N\equiv n_{N}+n_{\mathrm{A}}$, where $n_{N}\equiv\max\left(m,n\right)$
is the number of nominal modes, i.e. the number of modes explicitly
included in $T$, and $n_{\mathrm{A}}$ is the number of singular
values of $T$ not equal to 1. Modes $1$ to $n_{N}$ are associated
with the nominal modes, while modes $n_{N}+1$ to $N$ are associated
with full ancilla modes, by which we denote those modes that are added
throughout the whole transformation, not just as inputs or outputs
to match the number of input and output modes, as described in \textit{Step 1b}.
Each nominal mode $j$ has its own corresponding ancilla mode $m_{\mathrm{A}j}$
if the $j^{\mathrm{th}}$ singular value of $T$ differs from 1.

\paragraph{Step 4, finding associated quasiunitary matrices:}

We construct a matrix $S_{\mathrm{U}i}$ for each $U_{i}$ and similarly,
a matrix $S_{\mathrm{W}k}$ for each $W_{k}.$ $S_{\mathrm{U}i}$
and $S_{\mathrm{W}k}$ are defined as 
\begin{equation}
S_{\mathrm{U}i}\equiv\left(\begin{array}{cccc}
U_{i} & 0 & 0 & 0\\
0 & I_{n_{\mathrm{A}}} & 0 & 0\\
0 & 0 & U_{i}^{*} & 0\\
0 & 0 & 0 & I_{n_{\mathrm{A}}}
\end{array}\right)\label{eq:Sui}
\end{equation}

and 
\begin{equation}
S_{\mathrm{W}k}\equiv\left(\begin{array}{cccc}
W_{k} & 0 & 0 & 0\\
0 & I_{n_{\mathrm{A}}} & 0 & 0\\
0 & 0 & W_{k}^{*} & 0\\
0 & 0 & 0 & I_{n_{\mathrm{A}}}
\end{array}\right),\label{eq:Swi}
\end{equation}
respectively, with $U_{i}$ and $W_{k}$ being the $n_{N}\times n_{N}$
matrices from Step 2, $I_{n_{\mathrm{A}}}$ being the $n_{\mathrm{A}}\times n_{\mathrm{A}}$
identity matrix, and the $0$s being matrices of the appropriate size
filled with zeros. In addition, we construct a matrix $S_{\mathrm{D}j}$
for each $D_{j}$. For the special case that the $j^{\mathrm{th}}$
singular value $\sigma_{j}=1,$ $S_{\mathrm{D}j}$ is the $N\times N$
identity matrix and therefore not needed. Otherwise, if $\sigma_{j}\ne1$,
the matrix $S_{\mathrm{D}j}$ is the $N\times N$ identity matrix
with the elements corresponding to the intersection of rows and columns
$j,m_{\mathrm{A}j},j+N,m_{\mathrm{A}j}+N$ replaced by 
\begin{equation}
\left(\begin{array}{cccc}
\sigma_{j} & \sqrt{1-\sigma_{j}^{2}} & 0 & 0\\
-\sqrt{1-\sigma_{j}^{2}} & \sigma_{j} & 0 & 0\\
0 & 0 & \sigma_{j} & \sqrt{1-\sigma_{j}^{2}}\\
0 & 0 & -\sqrt{1-\sigma_{j}^{2}} & \sigma_{j}
\end{array}\right)\label{eq:ancexpr1}
\end{equation}
if $\sigma_{j}<1$, and 
\begin{equation}
\left(\begin{array}{cccc}
\sigma_{j} & 0 & 0 & \sqrt{\sigma_{j}^{2}-1}\\
0 & \sigma_{j} & \sqrt{\sigma_{j}^{2}-1} & 0\\
0 & \sqrt{\sigma_{j}^{2}-1} & \sigma_{j} & 0\\
\sqrt{\sigma_{j}^{2}-1} & 0 & 0 & \sigma_{j}
\end{array}\right)\label{eq:ancexpr2}
\end{equation}
if $\sigma_{j}>1$. This also allows dealing with transformations
that combine loss in some modes with gain in others, which previously
proposed methods did not accommodate. An example can be found in Appendix
\ref{subsec:Arbitrary-beam-splitter}.

\paragraph{Step 5, multiplication of quasiunitary matrices to obtain the overall
transformation:}

We obtain the overall enlarged transformation as
\begin{equation}
S_{\mathrm{total}}=\prod_{ijk}S_{\mathrm{U}i}S_{\mathrm{D}j}S_{\mathrm{W}k}.
\end{equation}
 A proof that $S_{\mathrm{total}}$ fulfills the quasiunitarity equation
(\ref{eq:quasiunitarity}) and contains $T$ as its upper left block
can be found in Appendix \ref{sec:AppProof}, and an example decomposition
is shown in Appendix \ref{subsec:Lossy-beam-splitter}.

\paragraph{Implementation of the decomposition in terms of simple building blocks}

The full decomposition $\mbox{\ensuremath{S_{\mathrm{total}}}=\ensuremath{\prod_{ijk}S_{\mathrm{U}i}S_{\mathrm{D}j}S_{\mathrm{W}k}}}$
provides a recipe for an implementation in terms of the simple building
blocks of phase shifters, beam splitters, and parametric amplifiers,
as each of the matrices in the decomposition directly corresponds
to such a building block. The factors $S_{\mathrm{U}i}$ and $S_{\mathrm{W}k}$
correspond to beam splitters and phase shifters involving the nominal
modes, i.e. the first $n_{N}$ modes. The factors $S_{\mathrm{D}j}$
that differ from the identity correspond to beam splitters and parametric
amplifiers, each involving one of the nominal modes and one of the
full ancilla modes.

\section{Discussion}

Section \ref{sec:Results} has shown how a full enlarged quantum optical
network can be mathematically represented and physically realized.
Now we are also in a position to answer the remaining questions from
the introduction. Contrary to conclusions of earlier works devoted
to setups with either loss or gain alone, any transformation is available.
The decomposition works for all linear networks as an input, since
a singular value decomposition can be performed for any complex matrix.
This means that in principle any transformation can be realized, even
if the practical implementation of arbitrary two-mode squeezing is
technically challenging \citep{Andersen2016}. 

The number of required
ancilla modes is tied to the dimensionality of $T$ if it is not square,
as well as to its singular values. A non-square $n\times m$ transformation
$T$ leads to $(m-n)$ output ancilla modes if $m>n$, or to $(n-m)$
input ancilla modes if $n>m$. In addition to these input or output
ancilla modes, full ancilla modes are introduced, and their number
is equal to the number of singular values of $T$ that are not equal
to 1. Each singular value below (above) 1 entails a beam splitter
operation (parametric amplification) with such an ancilla mode. For
the special case where $T$ is square and all of its singular values
are equal to 1, no ancilla modes are needed because $T$ is unitary,
and then the method can be reduced to the known unitary decompositions
(\citep{Reck1994} or \citep{Clements2016}). 
{\color{blue}Upper bounds on the number of elemental building blocks required when using the scheme depend on the dimensionality of $T$ in the following way: 
The maximum number of variable beam splitters needed to implement the unitary blocks $U$ and $W$ is\mbox{ $n(n-1)/2+m(m-1)/2$},
while the maximum number of phase shifters is \mbox{$n(n+1)/2+m(m+1)/2 $}.
Additionally, up to $\mathrm{min}(m,n)$ elements are required to implement $D$; these elements are either beam splitters or parametric amplifiers. 
Hence, the number of parametric amplifiers only scales linearly with the size of the transformation matrix. }

A unitary network followed by photon detection in the different modes
can be used to implement a projective measurement in a Hilbert space
with a dimensionality matching the unitary network. In the context
of generalized measurements, it is possible for a POVM to have a number
of measurement outcomes that is larger than the dimensionality of
the system. The Naimark dilation theorem guarantees that such a POVM
can be implemented as a projective measurement in an enlarged Hilbert
space \citep{Peres1990}. Our method can be used to find a Naimark
extension, which provides a suitable enlarged unitary transformation
for this projective measurement (see Appendix \ref{subsec:Simple-POVM}).

Another possible application of the method lies in the construction
of probabilistic optical quantum information protocols. Starting with
a general transformation matrix, by formulating the action of the
protocol as a mapping from a given set of input states to a set of
desired output states, a system of possibly nonlinear equations for
the elements of $T$ can be constructed. A solution of the system
of equations defines a network that performs the protocol, and the
method can then be used to find an implementation of that network
(for an example, see Appendix \ref{subsec:ProtocolDesign}). 

Although the decomposition always provides a full quantum optical
transformation with the dependence of the mean output fields on the
mean input fields as specified by the partial network $T$, the implementation
is not unique. This is already evident from the simplest nonunitary
`network', a single channel with loss or gain. As discussed in Ref.
\citealp{Leonhardt2003}, the same mean field could be achieved by
including excess gain and loss that compensate each other's effect
on the mean field, at the expense of a reduction in the purity of
the state. Given that this leaves the first moment of the field invariant
but changes higher order moments, it presents an opportunity to tailor
the higher order moments. It is an interesting question beyond the
scope of the present article whether the multi-mode control over first
moments of the field provided by the method could be extended to higher
order moments.

\section{Conclusion}

In summary, we have presented a way to describe and implement an arbitrary
linear optical transformation, which can have any size and does not
need to be complete in the sense that its modes fulfill bosonic commutation
relations. This is achieved by finding a transformation in an enlarged
space that includes the network of interest. The ancilla modes included
in the description enable rigorous quantum optical modeling of the
gain and losses in the network. In addition, a decomposition into
the basic building blocks of beam splitters, phase shifters, and parametric
amplifiers is obtained. This shows a way to implement the network
that could physically be realized with integrated optics. We have
discussed the role that the singular values of the transformation
matrix play with respect to the number and type of ancilla modes.
The method could prove useful for the implementation of POVMs, the
design of probabilistic optical quantum information protocols, and
more generally in any application that involves nonunitary networks. 

{\color{blue} We provide a MATLAB code for numerically implementing the full decomposition on GitHub, at https://github.com/NoraTischler/QuantOpt-linear-transformation-decomposition.}

\begin{acknowledgments} 
We wish to acknowledge discussions with Anton Zeilinger, Gabriel Molina-Terriza,
Geoff Pryde, Tim Ralph, Howard Wiseman, Michael Hall, and the group
of Stephen Barnett. Part of this work was supported by Australian
Research Council grant DP160101911, the Austrian Academy of Sciences
(\"{O}AW), the Austrian Science Fund (FWF) with SFB F40 (FOQUS). KS acknowledges
the support from the Foundation for Polish Science (project HEIMaT no. Homing/2016-1/8)
within the European Regional Development Fund. CR and KS also wish
to thank the Deutscher Akademischer Austauschdienst (PPP Poland) and
the Ministry of Science and Higher Education in Poland for support. 
\end{acknowledgments}

\appendix

\section{Proof that the product $S_{\mathrm{total}}=\prod_{ijk}S_{\mathrm{U}i}S_{\mathrm{D}j}S_{\mathrm{W}k}$
results in a quasiunitary matrix with T as its upper left block\label{sec:AppProof}}

First it should be noted that the individual $S$ matrices ($S_{\mathrm{U}i}$,
$S_{\mathrm{D}j}$, and $S_{\mathrm{W}k}$) fulfill Eq. (\ref{eq:quasiunitarity}).
The product of two matrices that fulfill Eq. (\ref{eq:quasiunitarity})
is another quasiunitary matrix, which can be seen as follows.

Let $A$ and $B$ fulfill Eq. (\ref{eq:quasiunitarity}). Then

\begin{align*}
M & =AB\\
MGM^{\dagger} & =\left(AB\right)G\left(AB\right)^{\dagger}\\
 & =A\left(BGB^{\dagger}\right)A^{\dagger}\\
 & =AGA^{\dagger}\\
 & =G.
\end{align*}
\\
Therefore, $S_{\mathrm{total}}$ is quasiunitary.

The second part of the proof is that the product of the individual
$S$ matrices has $T$ as its upper left block.

We have $T=UDW=\prod_{ijk}U_{i}D_{j}W_{k}$, and $S_{\mathrm{total}}=\prod_{ijk}S_{\mathrm{U}i}S_{\mathrm{D}j}S_{\mathrm{W}k}$.
Due to the block structure of the matrices $S_{\mathrm{U}i}$ and
$S_{\mathrm{W}k}$ , 

\[
\prod_{i}S_{\mathrm{U}i}=\left(\begin{array}{cccc}
\prod_{i}U_{i} & 0 & 0 & 0\\
0 & I_{n_{\mathrm{A}}} & 0 & 0\\
0 & 0 & \prod_{i'}U_{i'}^{*} & 0\\
0 & 0 & 0 & I_{n_{\mathrm{A}}}
\end{array}\right)
\]

and similarly 
\[
\prod_{k}S_{\mathrm{W}k}=\left(\begin{array}{cccc}
\prod_{k}W_{k} & 0 & 0 & 0\\
0 & I_{n_{\mathrm{A}}} & 0 & 0\\
0 & 0 & \prod_{k'}W_{k'}^{*} & 0\\
0 & 0 & 0 & I_{n_{\mathrm{A}}}
\end{array}\right).
\]

The components $S_{\mathrm{D}j}$ corresponding to $D_{j}$ do not
generally have the same structure. $S_{\mathrm{D}j}$ is the identity
matrix if the $j^{\mathrm{th}}$ singular value of $T$, $\sigma_{j}=1$.
Otherwise, if $\sigma_{j}\ne1$, $j$ and $m_{\mathrm{A}j}$ are the
mode numbers corresponding to the nominal mode and ancilla mode, respectively,
of the $j^{\mathrm{th}}$ singular value. Then, each matrix $S_{\mathrm{D}j}$
is the identity matrix with the elements corresponding to the intersection
of rows and columns $j,m_{\mathrm{A}j},j+N,m_{\mathrm{A}j}+N$ replaced
as given by expressions (\ref{eq:ancexpr1}) and (\ref{eq:ancexpr2}).

The fact that $S_{\mathrm{total}}=\prod_{ijk}S_{\mathrm{U}i}S_{\mathrm{D}j}S_{\mathrm{W}k}$
has $T=\prod_{ijk}U_{i}D_{j}W_{k}$ as its upper left block can be
shown by observing the structure of the matrix as the multiplication
is carried out. Let us consider the multiplication by starting from
the right-most matrix, sequentially multiplying from the left by the
other matrices as specified, and denoting the product after $x$ steps
as $S_{x}$. The rows of $S_{x}$ that deviate from those of the identity
matrix are of interest at different stages of the multiplication,
i.e. for different $x$. Let $x_{1}$ equal the number of matrices
in the decomposition of $W$. For $S_{x\mathrm{1}}=\prod_{k}S_{\mathrm{W}k}$
we have already seen that the upper left block of $S_{\mathrm{x1}}$
is the product of the upper left blocks of the components, and that
the only elements that deviate from the identity matrix are contained
in the blocks $\left(1:n_{N},1:n_{N}\right)$ and $\left(1+N:n_{N}+N,1+N:n_{N}+N\right)$
\footnote{By $\left(a:b,c:d\right)$ we denote the submatrix consisting of the
intersection of rows $a$ to $b$ and columns $c$ to $d$ of the
original matrix}. Now, as each $S_{\mathrm{D}j}$ is multiplied from the left, there
are at most two new rows of the resulting matrix that can deviate
from the identity: rows $m_{\mathrm{A}j}$ and $\left(m_{\mathrm{A}j}+N\right)$
when $\sigma_{j}\ne1$. Let $x_{2}$ lie between $x_{1}$ and the
number of matrices in the decomposition of $DW$. After each step,
the upper left block of $S_{\mathrm{x_{2}}}$ is the product of the
upper left blocks of the components because the elements $\left(m_{\mathrm{A}j},1:n_{\mathrm{N}}\right)$
and $\left(m_{\mathrm{A}j}+N,1:n_{\mathrm{N}}\right)$ of $S_{x_{2}\mathrm{-1}}$
are zero. This is essentially due to the fact that a unique ancilla
mode is assigned to each singular value different from 1. After having
multiplied through the individual S matrices corresponding to $DW$,
for $\prod_{i}S_{\mathrm{U}i}$ we again have the block structure
that guarantees that the upper left block of $S_{\mathrm{total}}$
is $T$.

\section{Examples}

We demonstrate the method on two examples. First, we discuss how the
lossy beam splitter with apparent nonlinearity can be constructed
with standard optical elements. We then apply the method to an arbitrary
$2\times2$ transformation, which may combine loss and gain in different
modes, to obtain an analytic decomposition. 

\subsection{Lossy beam splitter with apparent nonlinear action \label{subsec:Lossy-beam-splitter}}

\begin{figure}
\includegraphics{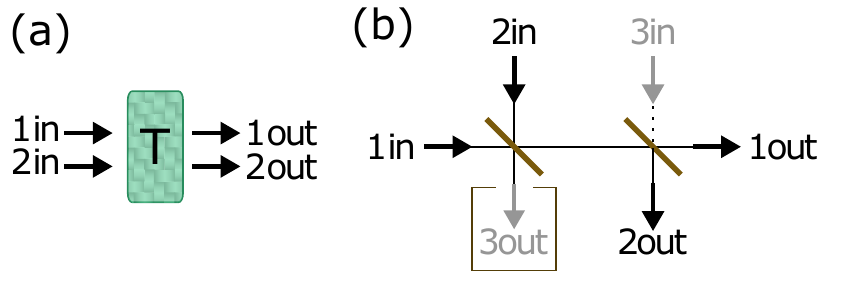}

\caption{Implementation of the lossy beam splitter with apparent nonlinear
loss. (a) One approach would be to implement the transformation directly
by a single device. The special transformation coefficients may be
achieved with a novel material, e.g. a metamaterial. (b) The decomposition
reveals a much simpler implementation consisting of two 50:50 beam
splitters (along with single-mode phase shifts omitted from the diagram),
and elucidates the simple role quantum interference plays with respect
to the `nonlinear loss'. \label{fig:example1-1}}
\end{figure}
Here, the method is applied to decompose the $2\times2$ transformation
$T=\frac{1}{2}\left(\begin{array}{cc}
1 & -1\\
-1 & 1
\end{array}\right)$ into simple building blocks. We begin with a singular value decomposition
of $T=UDW$, which gives $U=\frac{1}{\sqrt{2}}\left(\begin{array}{cc}
-1 & 1\\
1 & 1
\end{array}\right)$, $\mbox{D=\ensuremath{\left(\begin{array}{cc}
 1  &  0\\
 0  &  0 
\end{array}\right)}}$, and $W=\frac{1}{\sqrt{2}}\left(\begin{array}{cc}
-1 & 1\\
-1 & -1
\end{array}\right)$. Since $T$ is square, no augmentation of $U,$ $D,$ or $W$ is
required. Further decomposition provides $U=\left(\begin{array}{cc}
-1 & 0\\
0 & 1
\end{array}\right)\left(\begin{array}{cc}
\frac{1}{\sqrt{2}} & -\frac{1}{\sqrt{2}}\\
\frac{1}{\sqrt{2}} & \frac{1}{\sqrt{2}}
\end{array}\right)$ while $W$ is already a beam splitter, one of the basic building
blocks. The matrix $D$ does not need to be decomposed further due
to its simple form: the diagonal element $0$ in $D$ represents a
complete attenuation of a mode and constitutes the only singular value
different from $1$. One can thus proceed to identify the number of
ancilla modes $n_{A}=1$, so that $N=3$ and the dimensionality of
the corresponding $S$ matrix is $6\times6$ 
\begin{align*}
\left(\begin{array}{c}
\begin{array}{c}
\hat{a}_{1\mathrm{out}}\\
\hat{a}_{2\mathrm{out}}
\end{array}\\
\hat{a}_{3\mathrm{out}}\\
\hat{a}_{1\mathrm{out}}^{\dagger}\\
\hat{a}_{2\mathrm{out}}^{\dagger}\\
\hat{a}_{3\mathrm{out}}^{\dagger}
\end{array}\right) & =S\left(\begin{array}{c}
\begin{array}{c}
\hat{a}_{1\mathrm{in}}\\
\hat{a}_{2\mathrm{in}}
\end{array}\\
\hat{a}_{3\mathrm{in}}\\
\hat{a}_{1\mathrm{in}}^{\dagger}\\
\hat{a}_{2\mathrm{in}}^{\dagger}\\
\hat{a}_{3\mathrm{in}}^{\dagger}
\end{array}\right),
\end{align*}
with the nominal modes $\hat{a}_{1}$ and $\hat{a}_{2}$, and the
ancilla mode $\hat{a}_{3}$.

We continue to identify the $S_{U}$, $S_{D}$, and $S_{W}$ matrices
corresponding to individual operations based on Eqs. (\ref{eq:Sui})-(\ref{eq:ancexpr1}):
\begin{align*}
\prod_{i=1}^{2}S_{\mathrm{U}i} & =\left(\begin{array}{cccccc}
-1 & 0 & 0 & 0 & 0 & 0\\
0 & 1 & 0 & 0 & 0 & 0\\
0 & 0 & 1 & 0 & 0 & 0\\
0 & 0 & 0 & -1 & 0 & 0\\
0 & 0 & 0 & 0 & 1 & 0\\
0 & 0 & 0 & 0 & 0 & 1
\end{array}\right)\left(\begin{array}{cccccc}
\frac{1}{\sqrt{2}} & \frac{-1}{\sqrt{2}} & 0 & 0 & 0 & 0\\
\frac{1}{\sqrt{2}} & \frac{1}{\sqrt{2}} & 0 & 0 & 0 & 0\\
0 & 0 & 1 & 0 & 0 & 0\\
0 & 0 & 0 & \frac{1}{\sqrt{2}} & \frac{-1}{\sqrt{2}} & 0\\
0 & 0 & 0 & \frac{1}{\sqrt{2}} & \frac{1}{\sqrt{2}} & 0\\
0 & 0 & 0 & 0 & 0 & 1
\end{array}\right),
\end{align*}
\begin{eqnarray*}
S_{\mathrm{W}} & = & \left(\begin{array}{cccccc}
\frac{-1}{\sqrt{2}} & \frac{1}{\sqrt{2}} & 0 & 0 & 0 & 0\\
\frac{-1}{\sqrt{2}} & \frac{-1}{\sqrt{2}} & 0 & 0 & 0 & 0\\
0 & 0 & 1 & 0 & 0 & 0\\
0 & 0 & 0 & \frac{-1}{\sqrt{2}} & \frac{1}{\sqrt{2}} & 0\\
0 & 0 & 0 & \frac{-1}{\sqrt{2}} & \frac{-1}{\sqrt{2}} & 0\\
0 & 0 & 0 & 0 & 0 & 1
\end{array}\right),\\
S_{D} & = & \left(\begin{array}{cccccc}
1 & 0 & 0 & 0 & 0 & 0\\
0 & 0 & 1 & 0 & 0 & 0\\
0 & -1 & 0 & 0 & 0 & 0\\
0 & 0 & 0 & 1 & 0 & 0\\
0 & 0 & 0 & 0 & 0 & 1\\
0 & 0 & 0 & 0 & -1 & 0
\end{array}\right).
\end{eqnarray*}

\newpage
The total transformation matrix 
\begin{eqnarray*}
S_{\mathrm{total}} & = & \prod_{i}S_{\mathrm{U}i}S_{\mathrm{D}}S_{\mathrm{W}}=\left(\begin{array}{cccccc}
\frac{1}{2} & \frac{-1}{2} & \frac{1}{\sqrt{2}} & 0 & 0 & 0\\
\frac{-1}{2} & \frac{1}{2} & \frac{1}{\sqrt{2}} & 0 & 0 & 0\\
\frac{1}{\sqrt{2}} & \frac{1}{\sqrt{2}} & 0 & 0 & 0 & 0\\
0 & 0 & 0 & \frac{1}{2} & \frac{-1}{2} & \frac{1}{\sqrt{2}}\\
0 & 0 & 0 & \frac{-1}{2} & \frac{1}{2} & \frac{1}{\sqrt{2}}\\
0 & 0 & 0 & \frac{1}{\sqrt{2}} & \frac{1}{\sqrt{2}} & 0
\end{array}\right)
\end{eqnarray*}
indeed contains $T$ as its upper left block, and is consistent with
the scattering matrix given in Ref. \citealp{Roger2015}. Figure \ref{fig:example1-1}(b)
shows the setup after simplifications, such as rewriting the beam
splitter between modes 2 and 3 from $S_{D}$ in terms of a swap operation,
which means an exchange between the labels of the two modes. The setup
reveals that the apparent nonlinear loss is simply the result of photon
bunching due to two-photon quantum interference at the first beam
splitter; one of the output ports of the beam splitter is discarded,
which leads to either both or neither of the two photons emerging
in the nominal output modes 1 and 2.

\begin{figure}
\includegraphics[width=1\columnwidth]{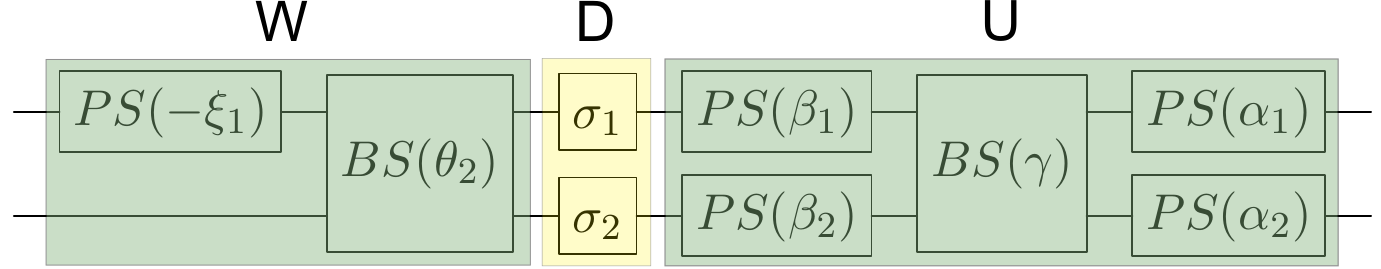}

\caption{Implementation of an arbitrary $2\times2$ transformation. The two
green rectangles represent the unitary components $W$ and $U$ of
the singular value decomposition. The diagonal part $D$, marked in
yellow, corresponds to single-mode modulations by $\sigma_{j}$, which
are realized by coupling to ancilla modes. \label{fig:example2}}
\end{figure}

\begin{widetext}
\subsection{ {\color{blue} General $\mathbf{2\times 2}$ linear transformation} \label{subsec:Arbitrary-beam-splitter}}

We now turn to a more general case of an arbitrary $2\times2$ linear
transformation matrix $T=\left(\begin{array}{cc}
t_{11} & t_{12}\\
t_{21} & t_{22}
\end{array}\right)$, with complex elements $t_{ij}=|t_{ij}|e^{i\varphi_{ij}}$, $\varphi_{ij}\in\mathbb{R}$.
Although the method always provides an easy way to obtain a decomposition
numerically, in this low-dimensional case, an analytical solution, depicted in Fig.\ \ref{fig:example2}, can also be found. We will represent the solution in terms of the following
matrices: rotations by a beam splitter of real coefficients
\begin{align*}
BS\left(\theta\right) & =\left(\begin{array}{cc}
\cos\theta & \sin\theta\\
-\sin\theta & \cos\theta
\end{array}\right)
\end{align*}
and single-mode phase shifts
\begin{align*}
PS_{1}\left(\theta\right) & =\left(\begin{array}{cc}
e^{i\theta} & 0\\
0 & 1
\end{array}\right),\ PS_{2}\left(\theta\right)=\left(\begin{array}{cc}
1 & 0\\
0 & e^{i\theta}
\end{array}\right).
\end{align*}

To solve this case analytically, one can transform the $T$ matrix
to a real form $T_{\mathrm{re}}$ through the following sequence of
operations:
\begin{enumerate}
\item cancel phases in the left column 
\begin{align*}
T\rightarrow T_{1} & =PS_{1}\left(-\varphi_{11}\right).PS_{2}\left(-\varphi_{21}\right).T=\left(\begin{array}{cc}
|t_{11}| & |t_{12}|e^{i\left(\varphi_{12}-\varphi_{11}\right)}\\
|t_{21}| & |t_{22}|e^{i\left(\varphi_{22}-\varphi_{21}\right)}
\end{array}\right),
\end{align*}
\\
\textcolor{black}{where matrix multiplication is indicated by ``$.$''
for clarity;}
\item rotate the matrix to null the bottom left component 
\begin{align*}
T_{1}\rightarrow T_{2} & =BS\left(\vartheta\right).T_{1}=\left(\begin{array}{cc}
\tilde{t}_{11} & \tilde{t}_{12}\\
0 & \tilde{t}_{22}
\end{array}\right),
\end{align*}
where $\vartheta=\arctan\left(\frac{|t_{21}|}{|t_{11}|}\right)$ and
\begin{align*}
\tilde{t}_{11} & =|t_{11}|\cos\vartheta+|t_{21}|\sin\vartheta,\\
\tilde{t}_{12} & =|t_{12}|\cos\vartheta e^{i\left(\varphi_{12}-\varphi_{11}\right)}+|t_{22}|\sin\vartheta e^{i\left(\varphi_{22}-\varphi_{21}\right)},\\
\tilde{t}_{22} & =-|t_{12}|\sin\vartheta e^{i\left(\varphi_{12}-\varphi_{11}\right)}+|t_{22}|\cos\vartheta e^{i\left(\varphi_{22}-\varphi_{21}\right)}.
\end{align*}
Note that $\tilde{t}_{11}$ is real and non-negative, which we will
emphasize below by explicitly writing \mbox{$\tilde{t}_{11}=|\tilde{t}_{11}|$};
\item cancel phases in the right column 
\begin{align*}
T_{2}\rightarrow T_{3}= & PS_{1}\left(-\xi_{1}\right).PS_{2}\left(-\xi_{2}\right).T_{2}=\left(\begin{array}{cc}
|\tilde{t}_{11}|e^{-i\xi_{1}} & |\tilde{t}_{12}|\\
0 & |\tilde{t}_{22}|
\end{array}\right),
\end{align*}
with $\xi_{j}=\arg\tilde{t}_{j2}$;
\item cancel the remaining phase in the left column
\begin{align*}
T_{3}\rightarrow T_{\mathrm{re}} & =T_{3}.PS_{1}\left(\xi_{1}\right)=\left(\begin{array}{cc}
|\tilde{t}_{11}| & |\tilde{t}_{12}|\\
0 & |\tilde{t}_{22}|
\end{array}\right).
\end{align*}
Finally, the transformed real matrix reads
\end{enumerate}
\begin{align}
T_{\mathrm{re}} & =PS_{1}\left(-\xi_{1}\right).PS_{2}\left(-\xi_{2}\right).BS\left(\vartheta\right).PS_{1}\left(-\varphi_{11}\right).PS_{2}\left(-\varphi_{21}\right).T.PS_{1}\left(\xi_{1}\right).\label{eq:Treal}
\end{align}
A singular value decomposition of the resulting real $2\times2$ matrix
is especially simple, with the unitary components given as two beam
splitter rotations. In this particular case we make use of the fact
that one of the components is $0$ and obtain 
\begin{align}
T_{\mathrm{re}} & =BS\left(\theta_{1}\right).D.BS\left(\theta_{2}\right),\label{eq:Treal_decomposition}
\end{align}
where
\begin{align*}
\theta_{j} & =\frac{\left(-1\right)^{j}}{2}\arg\left(q_{j}+2p_{j}i\right),\\
p_{j} & =|\tilde{t}_{1,3-j}\tilde{t}_{3-j,2}|,\\
q_{j} & =|\tilde{t}_{11}|^{2}-|\tilde{t}_{22}|^{2}+\left(-1\right)^{j-1}|\tilde{t}_{12}|^{2}.
\end{align*}
The matrix of singular values determines the required degree of attenuation
or amplification 
\begin{align*}
D & =\left(\begin{array}{cc}
\sigma_{1} & 0\\
0 & \sigma_{2}
\end{array}\right),\\
\sigma_{j} & =\sqrt{\frac{s+\left(-1\right)^{j-1}\sqrt{q_{j}^{2}+4p_{j}^{2}}}{2}},\\
s & =|\tilde{t}_{11}|^{2}+|\tilde{t}_{22}|^{2}+|\tilde{t}_{12}|^{2}.
\end{align*}
Finally, a combination of Eqs. (\ref{eq:Treal}) and (\ref{eq:Treal_decomposition})
yields the decomposition of the original matrix $T$
\[
T=\underbrace{PS_{2}\left(\varphi_{21}\right).PS_{1}\left(\varphi_{11}\right).BS\left(-\vartheta\right).PS_{2}\left(\xi_{2}\right).PS_{1}\left(\xi_{1}\right).BS\left(\theta_{1}\right)}_{U}.D.\underbrace{BS\left(\theta_{2}\right).PS_{1}\left(-\xi_{1}\right)}_{W}.
\]
Note that the matrix $U$ can be further simplified to 
\begin{align*}
U & =PS_{1}\left(\underbrace{\varphi_{11}+\xi_{1}+\frac{\alpha+\beta}{2}}_{\alpha_{1}}\right).PS_{2}\left(\underbrace{\varphi_{21}+\xi_{2}-\frac{\alpha+\beta}{2}}_{\alpha_{2}}\right).BS(\gamma).PS_{1}\left(\underbrace{\frac{\alpha-\beta}{2}}_{\beta_{1}}\right).PS_{2}\left(\underbrace{\frac{\beta-\alpha}{2}}_{\beta_{2}}\right),\\
\alpha & =\arg\left(\cos\vartheta\cos\theta_{1}+\sin\vartheta\sin\theta_{1}e^{i\left(\xi_{2}-\xi_{1}\right)}\right),\\
\beta & =\arg\left(\cos\vartheta\sin\theta_{1}-\sin\vartheta\cos\theta_{1}e^{i\left(\xi_{2}-\xi_{1}\right)}\right),\\
\gamma & ={\color{blue}{\color{black}\arccos}}\left(|\cos\vartheta\cos\theta_{1}+\sin\vartheta\sin\theta_{1}e^{i\left(\xi_{2}-\xi_{1}\right)}|\right).
\end{align*}

The construction of the $S$ network depends on the singular values
$\sigma_{1,2}$, and can be obtained from Eqs. (\ref{eq:Sui})-(\ref{eq:ancexpr2}).
The dimensionality of $S$ is at most $8\times8$, since there is
one ancilla mode per singular value $\neq1$. For a particular example,
the case of a transformation combining loss in mode $1$ ($\sigma_{1}<1$)
with gain in mode $2$ ($\sigma_{2}>1$), the submatrices read
\begin{align*}
S_{U} & =S_{PS_{2}\left(\alpha_{2}\right)}S_{PS_{1}\left(\alpha_{1}\right)}S_{BS\left(\gamma\right)}S_{PS_{2}\left(\beta_{2}\right)}S_{PS_{1}\left(\beta_{1}\right)}\\
 & =\left(\begin{array}{cccccccc}
e^{i\left(\alpha_{1}+\beta_{1}\right)}\cos\gamma & e^{i\left(\alpha_{1}+\beta_{2}\right)}\sin\gamma & 0 & 0\\
-e^{i\left(\alpha_{2}+\beta_{1}\right)}\sin\gamma & e^{i\left(\alpha_{2}+\beta_{2}\right)}\cos\gamma & 0 & 0\\
0 & 0 & 1 & 0\\
0 & 0 & 0 & 1\\
 &  &  &  & e^{-i\left(\alpha_{1}+\beta_{1}\right)}\cos\gamma & e^{-i\left(\alpha_{1}+\beta_{2}\right)}\sin\gamma & 0 & 0\\
 &  &  &  & -e^{-i\left(\alpha_{2}+\beta_{1}\right)}\sin\gamma & e^{-i\left(\alpha_{2}+\beta_{2}\right)}\cos\gamma & 0 & 0\\
 &  &  &  & 0 & 0 & 1 & 0\\
 &  &  &  & 0 & 0 & 0 & 1
\end{array}\right),
\end{align*}
\begin{align*}
S_{D1} & =\left(\begin{array}{cccccccc}
\sigma_{1} & 0 & \sqrt{1-\sigma_{1}^{2}} & 0\\
0 & 1 & 0 & 0\\
-\sqrt{1-\sigma_{1}^{2}} & 0 & \sigma_{1} & 0\\
0 & 0 & 0 & 1\\
 &  &  &  & \sigma_{1} & 0 & \sqrt{1-\sigma_{1}^{2}} & 0\\
 &  &  &  & 0 & 1 & 0 & 0\\
 &  &  &  & -\sqrt{1-\sigma_{1}^{2}} & 0 & \sigma_{1} & 0\\
 &  &  &  & 0 & 0 & 0 & 1
\end{array}\right),
\end{align*}
\begin{align*}
S_{D2} & =\left(\begin{array}{cccccccc}
1 & 0 & 0 & 0 & 0 & 0 & 0 & 0\\
0 & \sigma_{2} & 0 & 0 & 0 & 0 & 0 & \sqrt{\sigma_{2}^{2}-1}\\
0 & 0 & 1 & 0 & 0 & 0 & 0 & 0\\
0 & 0 & 0 & \sigma_{2} & 0 & \sqrt{\sigma_{2}^{2}-1} & 0 & 0\\
0 & 0 & 0 & 0 & 1 & 0 & 0 & 0\\
0 & 0 & 0 & \sqrt{\sigma_{2}^{2}-1} & 0 & \sigma_{2} & 0 & 0\\
0 & 0 & 0 & 0 & 0 & 0 & 1 & 0\\
0 & \sqrt{\sigma_{2}^{2}-1} & 0 & 0 & 0 & 0 & 0 & \sigma_{2}
\end{array}\right),
\end{align*}
\begin{align*}
S_{W} & =S_{BS\left(\theta_{2}\right)}S_{PS_{1}\left(-\xi_{1}\right)}\\
 & =\left(\begin{array}{cccccccc}
e^{-i\xi_{1}}\cos\theta_{2} & \sin\theta_{2} & 0 & 0\\
-e^{-i\xi_{1}}\sin\theta_{2} & \cos\theta_{2} & 0 & 0\\
0 & 0 & 1 & 0\\
0 & 0 & 0 & 1\\
 &  &  &  & e^{i\xi_{1}}\cos\theta_{2} & \sin\theta_{2} & 0 & 0\\
 &  &  &  & -e^{i\xi_{1}}\sin\theta_{2} & \cos\theta_{2} & 0 & 0\\
 &  &  &  & 0 & 0 & 1 & 0\\
 &  &  &  & 0 & 0 & 0 & 1
\end{array}\right),
\end{align*}
where the empty blocks should be filled with zeros. \end{widetext}

\section{Applications}

In this appendix we outline two applications in which the method can
be used: finding Naimark extensions for POVMs and the design of probabilistic
optical quantum information protocols.

\subsection{POVMs \label{subsec:Simple-POVM}}

A POVM is determined by a set of positive semidefinite operators $\left\{ E_{i}\right\} _{i=1}^{m}$,
which sum to identity $\mbox{\ensuremath{\sum_{i=1}^{m}E_{i}}=\ensuremath{I_{n}}}$
and represent generalized measurements in an $n$-dimensional Hilbert
space \citep{Nielsen}. Here, $I_{n}$ denotes the $n$-dimensional
identity matrix. An active field of research has been
focused on the physical implementation of POVMs \citep{He2006,Tabia2012,Bian2015,Bent2015,Sosa2017}.
One of the strategies is based on Naimark's dilation theorem. According
to the theorem, any POVM can be realized as a projective measurement
in an enlarged Hilbert space $\mathcal{H}$ \citep{Peres1990}. However,
the theorem does not itself provide a general recipe to find the extension
to $\mathcal{H}$, called Naimark's extension. 

To see how our method can be exploited to find Naimark extensions,
let us focus on the important case of rank-one POVMs. The operators
forming rank-one POVMs correspond to projectors $E_{i}=|\phi^{(i)}\rangle\langle\phi^{(i)}|$
on, in general, nonorthogonal vectors $|\phi^{(i)}\rangle$ in the
original Hilbert space. A Naimark extension can be found by augmenting
the vectors $|\phi^{(i)}\rangle$ to the size\textbf{ $m$} so that
they become orthogonal in $\mathcal{H}$. For this purpose, let us
define a rectangular $n\times m$ matrix with columns given by the
$n$-dimensional vectors $|\phi^{(i)}\rangle$:

\begin{align*}
T & =\left(\begin{array}{ccc}
\phi_{1}^{(1)} & \ldots & \phi_{1}^{(m)}\\
\vdots &  & \vdots\\
\phi_{n}^{(1)} & \ldots & \phi_{n}^{(m)}
\end{array}\right),
\end{align*}
such that $TT^{\dagger}=I$. Here, $\phi_{j}^{(i)}$ stand for elements
of $|\phi^{(i)}\rangle$. A singular value decomposition of $T=UDW$
provides a unitary $n\times n$ matrix $U$, an $n\times m$ matrix
$D$, and a unitary $m\times m$ matrix $W$. Note that since $TT^{\dagger}=I$,
all the singular values of $T$ are equal to $1$. This means that
the dimensionality of the Naimark extension found with the method
is $m$, and the number of ancilla output modes is $m-n$. Next, let
us pad the matrices of smaller dimensionalities with elements of the
identity matrix, in accordance with Step 1b of the method. As a result,
we obtain the enlarged $m\times m$ matrices: 
\begin{align*}
U & \rightarrow\left(\begin{array}{cc}
U & 0\\
0 & I_{m-n}
\end{array}\right),\\
D & \rightarrow I_{m},
\end{align*}
and $W$ does not require any modification. The product $UDW=UW$
is unitary and becomes an $m$-dimensional Naimark extension of $T$,
which can be directly decomposed into building blocks with methods
of Reck \textit{et al.} \citep{Reck1994} or Clements \textit{et al.}
\citep{Clements2016}. This procedure allows designing a network for
an arbitrary rank-one POVM. 

\subsection{Design of probabilistic protocols\label{subsec:ProtocolDesign}}

Here, we demonstrate how the method can be used in the design of probabilistic
optical quantum logic gates. We illustrate the design on the example
of the 2-qubit controlled-Z gate, and show a systematic way to find
the setup presented in Ref. \citealp{Hofmann2002}. A 2-qubit controlled-Z
gate can be implemented with two photons and four optical modes. The
control qubit is encoded by one photon within the first two modes
(called the control modes), while the target qubit is encoded by another
photon in the last two modes (the target modes). The goal is to construct
a transformation using passive optical elements, such that it implements
a controlled phase flip, given that both the input and output states
fulfill the condition that there is one photon in the control modes
and one photon in the target modes. 

Our starting point is the desired effect on two-photon states: for
the four different input states below and only considering outputs
according to the postselection condition of having one photon in a
control mode and the other photon in a target mode, we want the circuit
to output the following states:
\begin{eqnarray}
\hat{a}_{cH\mathrm{in}}\hat{a}_{tH\mathrm{in}} & \rightarrow & -k\hat{a}_{cH\mathrm{out}}\hat{a}_{tH\mathrm{out}}\nonumber \\
\hat{a}_{cH\mathrm{in}}\hat{a}_{tV\mathrm{in}} & \rightarrow & k\hat{a}_{cH\mathrm{out}}\hat{a}_{tV\mathrm{out}}\nonumber \\
\hat{a}_{cV\mathrm{in}}\hat{a}_{tH\mathrm{in}} & \rightarrow & k\hat{a}_{cV\mathrm{out}}\hat{a}_{tH\mathrm{out}}\nonumber \\
\hat{a}_{cV\mathrm{in}}\hat{a}_{tV\mathrm{in}} & \rightarrow & k\hat{a}_{cV\mathrm{out}}\hat{a}_{tV\mathrm{out}},\label{eq:conditions}
\end{eqnarray}
where the four modes are denoted $cH$, $cV$, $tH$, $tV$, after
horizontal and vertical polarization in the control and target modes.
The real constant $k\in(0,1]$ allows for the possibility of the protocol
being probabilistic, with a success rate of $k^{2}$. The above transformations
involve four input and output modes, so the transformation we seek
has the general form 
\[
T=\left(\begin{array}{cccc}
t_{11} & t_{12} & t_{13} & t_{14}\\
t_{21} & t_{22} & t_{23} & t_{24}\\
t_{31} & t_{32} & t_{33} & t_{34}\\
t_{41} & t_{42} & t_{43} & t_{44}
\end{array}\right).
\]
Since we assume that the setup will be passive, we know that $S_{\mathrm{total}}$
will be block-diagonal and can be written as 
\[
S_{\mathrm{total}}=\left(\begin{array}{cc}
A & 0\\
0 & A^{*}
\end{array}\right),
\]
where $A$ is a unitary matrix that contains $T$ as its upper left
block, relating annihilation operators as follows:
\[
\left(\begin{array}{c}
\hat{a}_{cH\mathrm{out}}\\
\hat{a}_{cV\mathrm{out}}\\
\hat{a}_{tH\mathrm{out}}\\
\hat{a}_{tV\mathrm{out}}\\
\vdots
\end{array}\right)=\left(\begin{array}{cc}
T & A_{12}\\
A_{21} & A_{22}
\end{array}\right)\left(\begin{array}{c}
\hat{a}_{cH\mathrm{in}}\\
\hat{a}_{cV\mathrm{in}}\\
\hat{a}_{tH\mathrm{in}}\\
\hat{a}_{tV\mathrm{in}}\\
\vdots
\end{array}\right).
\]

Based on the constraints of Eq. (\ref{eq:conditions}), the elements
of $T$ need to be determined. To do this, it is useful to write the
annihilation operators of the input modes in terms of those of the
output modes. The unitary matrix $A$ can simply be inverted to write
the input modes in terms of the output modes, and we obtain 
\begin{equation}
\left(\begin{array}{c}
\hat{a}_{cH\mathrm{in}}\\
\hat{a}_{cV\mathrm{in}}\\
\hat{a}_{tH\mathrm{in}}\\
\hat{a}_{tV\mathrm{in}}\\
\vdots
\end{array}\right)=\left(\begin{array}{cc}
T^{\dagger} & A_{21}^{\dagger}\\
A_{12}^{\dagger} & A_{22}^{\dagger}
\end{array}\right)\left(\begin{array}{c}
\hat{a}_{cH\mathrm{out}}\\
\hat{a}_{cV\mathrm{out}}\\
\hat{a}_{tH\mathrm{out}}\\
\hat{a}_{tV\mathrm{out}}\\
\vdots
\end{array}\right).\label{eq:invertedT}
\end{equation}

Using \foreignlanguage{american}{Eq}. (\ref{eq:conditions}) together
with Eq. (\ref{eq:invertedT}) provides a set of nonlinear equations,
of which one solution is 
\[
T=\left(\begin{array}{cccc}
t_{11} & 0 & t_{13} & 0\\
0 & t_{11} & 0 & 0\\
t_{31} & 0 & -\frac{t_{13}t_{31}}{2t_{11}} & 0\\
0 & 0 & 0 & -\frac{t_{13}t_{31}}{2t_{11}}
\end{array}\right),k=-\frac{1}{2}t_{13}t_{31}.
\]
There are three free parameters, $t_{11}$, $t_{13}$, $t_{31}$,
and the success probability of the protocol, $k^{2}$, depends on
two of these parameters. Moreover, the singular values of $T$ depend
on the parameters. We need all the singular values to be $\leq1$,
so that the circuit is a passive network, but would like as many of
the values as possible to be $1$, so that the number of ancilla modes
is minimized. A suitable choice of parameters is $t_{11}=\sqrt{\frac{1}{3}}$,
$t_{13}=t_{31}=\sqrt{\frac{2}{3}}$. This results in the success probability
of the protocol $k^{2}=\frac{1}{9}$, and the singular values $\left(1,1,\sqrt{\frac{1}{3}},\sqrt{\frac{1}{3}}\right),$
which show that two ancilla modes are required. From here, the decomposition
method can be used to find the physical realization of the matrix
\[
T=\left(\begin{array}{cccc}
\sqrt{\frac{1}{3}} & 0 & \sqrt{\frac{2}{3}} & 0\\
0 & \sqrt{\frac{1}{3}} & 0 & 0\\
\sqrt{\frac{2}{3}} & 0 & -\sqrt{\frac{1}{3}} & 0\\
0 & 0 & 0 & -\sqrt{\frac{1}{3}}
\end{array}\right),
\]
which finally provides the scheme of Ref. \citealp{Hofmann2002}.

\bibliographystyle{revtex41}
\bibliography{biblio_decomp_final}

\begin{thebibliography}{33}%
\makeatletter
\providecommand \@ifxundefined [1]{%
 \@ifx{#1\undefined}
}%
\providecommand \@ifnum [1]{%
 \ifnum #1\expandafter \@firstoftwo
 \else \expandafter \@secondoftwo
 \fi
}%
\providecommand \@ifx [1]{%
 \ifx #1\expandafter \@firstoftwo
 \else \expandafter \@secondoftwo
 \fi
}%
\providecommand \natexlab [1]{#1}%
\providecommand \emph  [1]{``#1''}%
\providecommand \bibnamefont  [1]{#1}%
\providecommand \bibfnamefont [1]{#1}%
\providecommand \citenamefont [1]{#1}%
\providecommand \href@noop [0]{\@secondoftwo}%
\providecommand \href [0]{\begingroup \@sanitize@url \@href}%
\providecommand \@href[1]{\@@startlink{#1}\@@href}%
\providecommand \@@href[1]{\endgroup#1\@@endlink}%
\providecommand \@sanitize@url [0]{\catcode `\\12\catcode `\$12\catcode
  `\&12\catcode `\#12\catcode `\^12\catcode `\_12\catcode `\%12\relax}%
\providecommand \@@startlink[1]{}%
\providecommand \@@endlink[0]{}%
\providecommand \url  [0]{\begingroup\@sanitize@url \@url }%
\providecommand \@url [1]{\endgroup\@href {#1}{\urlprefix }}%
\providecommand \urlprefix  [0]{URL }%
\providecommand \Eprint [0]{\href }%
\providecommand \doibase [0]{http://dx.doi.org/}%
\providecommand \selectlanguage [0]{\@gobble}%
\providecommand \bibinfo  [0]{\@secondoftwo}%
\providecommand \bibfield  [0]{\@secondoftwo}%
\providecommand \translation [1]{[#1]}%
\providecommand \BibitemOpen [0]{}%
\providecommand \bibitemStop [0]{}%
\providecommand \bibitemNoStop [0]{.\EOS\space}%
\providecommand \EOS [0]{\spacefactor3000\relax}%
\providecommand \BibitemShut  [1]{\csname bibitem#1\endcsname}%
\let\auto@bib@innerbib\@empty
\bibitem [{\citenamefont {Reck}\ \emph {et~al.}(1994)\citenamefont {Reck},
  \citenamefont {Zeilinger}, \citenamefont {Bernstein},\ and\ \citenamefont
  {Bertani}}]{Reck1994}%
  \BibitemOpen
  \bibfield  {author} {\bibinfo {author} {\bibfnamefont {M.}~\bibnamefont
  {Reck}}, \bibinfo {author} {\bibfnamefont {A.}~\bibnamefont {Zeilinger}},
  \bibinfo {author} {\bibfnamefont {H.~J.}\ \bibnamefont {Bernstein}}, \ and\
  \bibinfo {author} {\bibfnamefont {P.}~\bibnamefont {Bertani}},\ }\bibfield
  {title} {\emph {\bibinfo {title} {Experimental realization of any discrete
  unitary operator},}\ }\href@noop {} {\bibfield  {journal} {\bibinfo
  {journal} {Phys. Rev. Lett.}\ }\textbf {\bibinfo {volume} {73}},\ \bibinfo
  {pages} {58} (\bibinfo {year} {1994})}\BibitemShut {NoStop}%
\bibitem [{Note1()}]{Note1}%
  \BibitemOpen
  \bibinfo {note} {By the terms `linear transformation' and `linear network' we
  refer to transformations for which the expectation values of the fields are
  related by a linear transformation between the input and output modes and the
  annihilation operators of the output modes have the same linear dependence on
  the input annihilation operators.}\BibitemShut {Stop}%
\bibitem [{\citenamefont {Clements}\ \emph {et~al.}(2016)\citenamefont
  {Clements}, \citenamefont {Humphreys}, \citenamefont {Metcalf}, \citenamefont
  {Kolthammer},\ and\ \citenamefont {Walsmley}}]{Clements2016}%
  \BibitemOpen
  \bibfield  {author} {\bibinfo {author} {\bibfnamefont {W.~R.}\ \bibnamefont
  {Clements}}, \bibinfo {author} {\bibfnamefont {P.~C.}\ \bibnamefont
  {Humphreys}}, \bibinfo {author} {\bibfnamefont {B.~J.}\ \bibnamefont
  {Metcalf}}, \bibinfo {author} {\bibfnamefont {W.~S.}\ \bibnamefont
  {Kolthammer}}, \ and\ \bibinfo {author} {\bibfnamefont {I.~A.}\ \bibnamefont
  {Walsmley}},\ }\bibfield  {title} {\emph {\bibinfo {title} {Optimal design
  for universal multiport interferometers},}\ }\href@noop {} {\bibfield
  {journal} {\bibinfo  {journal} {Optica}\ }\textbf {\bibinfo {volume} {3}},\
  \bibinfo {pages} {1460} (\bibinfo {year} {2016})}\BibitemShut {NoStop}%
\bibitem [{\citenamefont {Barnett}\ \emph {et~al.}(1998)\citenamefont
  {Barnett}, \citenamefont {Jeffers}, \citenamefont {Gatti},\ and\
  \citenamefont {Loudon}}]{Barnett1998}%
  \BibitemOpen
  \bibfield  {author} {\bibinfo {author} {\bibfnamefont {S.~M.}\ \bibnamefont
  {Barnett}}, \bibinfo {author} {\bibfnamefont {J.}~\bibnamefont {Jeffers}},
  \bibinfo {author} {\bibfnamefont {A.}~\bibnamefont {Gatti}}, \ and\ \bibinfo
  {author} {\bibfnamefont {R.}~\bibnamefont {Loudon}},\ }\bibfield  {title}
  {\emph {\bibinfo {title} {Quantum optics of lossy beam splitters},}\
  }\href@noop {} {\bibfield  {journal} {\bibinfo  {journal} {Phys. Rev. A}\
  }\textbf {\bibinfo {volume} {57}},\ \bibinfo {pages} {2134} (\bibinfo {year}
  {1998})}\BibitemShut {NoStop}%
\bibitem [{\citenamefont {Kn{\"o}ll}\ \emph {et~al.}(1999)\citenamefont
  {Kn{\"o}ll}, \citenamefont {Scheel}, \citenamefont {Schmidt}, \citenamefont
  {Welsch},\ and\ \citenamefont {Chizhov}}]{Knoell1999}%
  \BibitemOpen
  \bibfield  {author} {\bibinfo {author} {\bibfnamefont {L.}~\bibnamefont
  {Kn{\"o}ll}}, \bibinfo {author} {\bibfnamefont {S.}~\bibnamefont {Scheel}},
  \bibinfo {author} {\bibfnamefont {E.}~\bibnamefont {Schmidt}}, \bibinfo
  {author} {\bibfnamefont {D.-G.}\ \bibnamefont {Welsch}}, \ and\ \bibinfo
  {author} {\bibfnamefont {A.~V.}\ \bibnamefont {Chizhov}},\ }\bibfield
  {title} {\emph {\bibinfo {title} {Quantum-state transformation by dispersive
  and absorbing four-port devices},}\ }\href@noop {} {\bibfield  {journal}
  {\bibinfo  {journal} {Phys. Rev. A}\ }\textbf {\bibinfo {volume} {59}},\
  \bibinfo {pages} {4716} (\bibinfo {year} {1999})}\BibitemShut {NoStop}%
\bibitem [{\citenamefont {Jeffers}(2000)}]{Jeffers2000}%
  \BibitemOpen
  \bibfield  {author} {\bibinfo {author} {\bibfnamefont {J.}~\bibnamefont
  {Jeffers}},\ }\bibfield  {title} {\emph {\bibinfo {title} {Interference and
  the lossless lossy beam splitter},}\ }\href@noop {} {\bibfield  {journal}
  {\bibinfo  {journal} {J. Mod. Opt.}\ }\textbf {\bibinfo {volume} {47}},\
  \bibinfo {pages} {1819} (\bibinfo {year} {2000})}\BibitemShut {NoStop}%
\bibitem [{\citenamefont {Scheel}\ \emph {et~al.}(2000)\citenamefont {Scheel},
  \citenamefont {Kn\"oll}, \citenamefont {Opatrn\'y},\ and\ \citenamefont
  {Welsch}}]{Scheel2000}%
  \BibitemOpen
  \bibfield  {author} {\bibinfo {author} {\bibfnamefont {S.}~\bibnamefont
  {Scheel}}, \bibinfo {author} {\bibfnamefont {L.}~\bibnamefont {Kn\"oll}},
  \bibinfo {author} {\bibfnamefont {T.}~\bibnamefont {Opatrn\'y}}, \ and\
  \bibinfo {author} {\bibfnamefont {D.-G.}\ \bibnamefont {Welsch}},\ }\bibfield
   {title} {\emph {\bibinfo {title} {Entanglement transformation at absorbing
  and amplifying four-port devices},}\ }\href {\doibase
  10.1103/PhysRevA.62.043803} {\bibfield  {journal} {\bibinfo  {journal} {Phys.
  Rev. A}\ }\textbf {\bibinfo {volume} {62}},\ \bibinfo {pages} {043803}
  (\bibinfo {year} {2000})}\BibitemShut {NoStop}%
\bibitem [{\citenamefont {Lee}\ \emph {et~al.}(2012)\citenamefont {Lee},
  \citenamefont {Tame}, \citenamefont {Lim},\ and\ \citenamefont
  {Lee}}]{Lee2012}%
  \BibitemOpen
  \bibfield  {author} {\bibinfo {author} {\bibfnamefont {C.}~\bibnamefont
  {Lee}}, \bibinfo {author} {\bibfnamefont {M.}~\bibnamefont {Tame}}, \bibinfo
  {author} {\bibfnamefont {J.}~\bibnamefont {Lim}}, \ and\ \bibinfo {author}
  {\bibfnamefont {J.}~\bibnamefont {Lee}},\ }\bibfield  {title} {\emph
  {\bibinfo {title} {Quantum plasmonics with a metal nanoparticle array},}\
  }\href {\doibase 10.1103/PhysRevA.85.063823} {\bibfield  {journal} {\bibinfo
  {journal} {Phys. Rev. A}\ }\textbf {\bibinfo {volume} {85}},\ \bibinfo
  {pages} {063823} (\bibinfo {year} {2012})}\BibitemShut {NoStop}%
\bibitem [{\citenamefont {Miller}(2013)}]{Miller2013}%
  \BibitemOpen
  \bibfield  {author} {\bibinfo {author} {\bibfnamefont {D.~A.~B.}\
  \bibnamefont {Miller}},\ }\bibfield  {title} {\emph {\bibinfo {title}
  {Self-configuring universal linear optical component},}\ }\href@noop {}
  {\bibfield  {journal} {\bibinfo  {journal} {Photon. Res.}\ }\textbf {\bibinfo
  {volume} {1}},\ \bibinfo {pages} {1} (\bibinfo {year} {2013})}\BibitemShut
  {NoStop}%
\bibitem [{\citenamefont {Miller}(2012)}]{Miller2012}%
  \BibitemOpen
  \bibfield  {author} {\bibinfo {author} {\bibfnamefont {D.~A.~B.}\
  \bibnamefont {Miller}},\ }\bibfield  {title} {\emph {\bibinfo {title} {All
  linear optical devices are mode converters},}\ }\href@noop {} {\bibfield
  {journal} {\bibinfo  {journal} {Opt. Express}\ }\textbf {\bibinfo {volume}
  {20}},\ \bibinfo {pages} {23985} (\bibinfo {year} {2012})}\BibitemShut
  {NoStop}%
\bibitem [{\citenamefont {Dutta~Gupta}\ and\ \citenamefont
  {Agarwal}(2014)}]{Gupta2014}%
  \BibitemOpen
  \bibfield  {author} {\bibinfo {author} {\bibfnamefont {S.}~\bibnamefont
  {Dutta~Gupta}}\ and\ \bibinfo {author} {\bibfnamefont {G.~S.}\ \bibnamefont
  {Agarwal}},\ }\bibfield  {title} {\emph {\bibinfo {title} {Two-photon quantum
  interference in plasmonics: theory and applications},}\ }\href@noop {}
  {\bibfield  {journal} {\bibinfo  {journal} {Opt. Lett.}\ }\textbf {\bibinfo
  {volume} {39}},\ \bibinfo {pages} {390} (\bibinfo {year} {2014})}\BibitemShut
  {NoStop}%
\bibitem [{\citenamefont {Roger}\ \emph {et~al.}(2015)\citenamefont {Roger},
  \citenamefont {Vezzoli}, \citenamefont {Bolduc}, \citenamefont {Valente},
  \citenamefont {Heitz}, \citenamefont {Jeffers}, \citenamefont {Soci},
  \citenamefont {Leach}, \citenamefont {Couteau}, \citenamefont {Zheludev},\
  and\ \citenamefont {Faccio}}]{Roger2015}%
  \BibitemOpen
  \bibfield  {author} {\bibinfo {author} {\bibfnamefont {T.}~\bibnamefont
  {Roger}}, \bibinfo {author} {\bibfnamefont {S.}~\bibnamefont {Vezzoli}},
  \bibinfo {author} {\bibfnamefont {E.}~\bibnamefont {Bolduc}}, \bibinfo
  {author} {\bibfnamefont {J.}~\bibnamefont {Valente}}, \bibinfo {author}
  {\bibfnamefont {J.~J.~F.}\ \bibnamefont {Heitz}}, \bibinfo {author}
  {\bibfnamefont {J.}~\bibnamefont {Jeffers}}, \bibinfo {author} {\bibfnamefont
  {C.}~\bibnamefont {Soci}}, \bibinfo {author} {\bibfnamefont {J.}~\bibnamefont
  {Leach}}, \bibinfo {author} {\bibfnamefont {C.}~\bibnamefont {Couteau}},
  \bibinfo {author} {\bibfnamefont {N.~I.}\ \bibnamefont {Zheludev}}, \ and\
  \bibinfo {author} {\bibfnamefont {D.}~\bibnamefont {Faccio}},\ }\bibfield
  {title} {\emph {\bibinfo {title} {Coherent perfect absorption in deeply
  subwavelength films in the single-photon regime},}\ }\href@noop {} {\bibfield
   {journal} {\bibinfo  {journal} {Nat. Commun.}\ }\textbf {\bibinfo {volume}
  {6}},\ \bibinfo {pages} {7031} (\bibinfo {year} {2015})}\BibitemShut
  {NoStop}%
\bibitem [{\citenamefont {Roger}\ \emph {et~al.}(2016)\citenamefont {Roger},
  \citenamefont {Restuccia}, \citenamefont {Lyons}, \citenamefont {Giovannini},
  \citenamefont {Romero}, \citenamefont {Jeffers}, \citenamefont {Padgett},\
  and\ \citenamefont {Faccio}}]{Roger2016}%
  \BibitemOpen
  \bibfield  {author} {\bibinfo {author} {\bibfnamefont {T.}~\bibnamefont
  {Roger}}, \bibinfo {author} {\bibfnamefont {S.}~\bibnamefont {Restuccia}},
  \bibinfo {author} {\bibfnamefont {A.}~\bibnamefont {Lyons}}, \bibinfo
  {author} {\bibfnamefont {D.}~\bibnamefont {Giovannini}}, \bibinfo {author}
  {\bibfnamefont {J.}~\bibnamefont {Romero}}, \bibinfo {author} {\bibfnamefont
  {J.}~\bibnamefont {Jeffers}}, \bibinfo {author} {\bibfnamefont
  {M.}~\bibnamefont {Padgett}}, \ and\ \bibinfo {author} {\bibfnamefont
  {D.}~\bibnamefont {Faccio}},\ }\bibfield  {title} {\emph {\bibinfo {title}
  {Coherent absorption of n00n states},}\ }\href {\doibase
  10.1103/PhysRevLett.117.023601} {\bibfield  {journal} {\bibinfo  {journal}
  {Phys. Rev. Lett.}\ }\textbf {\bibinfo {volume} {117}},\ \bibinfo {pages}
  {023601} (\bibinfo {year} {2016})}\BibitemShut {NoStop}%
\bibitem [{\citenamefont {Uppu}\ \emph {et~al.}(2016)\citenamefont {Uppu},
  \citenamefont {Wolterink}, \citenamefont {Tentrup},\ and\ \citenamefont
  {Pinkse}}]{Uppu2016}%
  \BibitemOpen
  \bibfield  {author} {\bibinfo {author} {\bibfnamefont {R.}~\bibnamefont
  {Uppu}}, \bibinfo {author} {\bibfnamefont {T.~A.~W.}\ \bibnamefont
  {Wolterink}}, \bibinfo {author} {\bibfnamefont {T.~B.~H.}\ \bibnamefont
  {Tentrup}}, \ and\ \bibinfo {author} {\bibfnamefont {P.~W.~H.}\ \bibnamefont
  {Pinkse}},\ }\bibfield  {title} {\emph {\bibinfo {title} {Quantum optics of
  lossy asymmetric beam splitters},}\ }\href@noop {} {\bibfield  {journal}
  {\bibinfo  {journal} {Opt. Express}\ }\textbf {\bibinfo {volume} {24}},\
  \bibinfo {pages} {16440} (\bibinfo {year} {2016})}\BibitemShut {NoStop}%
\bibitem [{\citenamefont {Di~Martino}\ \emph {et~al.}(2014)\citenamefont
  {Di~Martino}, \citenamefont {Sonnefraud}, \citenamefont {Tame}, \citenamefont
  {K\'{e}na-Cohen}, \citenamefont {Dieleman}, \citenamefont {\"{O}zdemir},
  \citenamefont {Kim},\ and\ \citenamefont {Maier}}]{DiMartino2014}%
  \BibitemOpen
  \bibfield  {author} {\bibinfo {author} {\bibfnamefont {G.}~\bibnamefont
  {Di~Martino}}, \bibinfo {author} {\bibfnamefont {Y.}~\bibnamefont
  {Sonnefraud}}, \bibinfo {author} {\bibfnamefont {M.~S.}\ \bibnamefont
  {Tame}}, \bibinfo {author} {\bibfnamefont {S.}~\bibnamefont
  {K\'{e}na-Cohen}}, \bibinfo {author} {\bibfnamefont {F.}~\bibnamefont
  {Dieleman}}, \bibinfo {author} {\bibfnamefont {{\c{S}}.~K.}\ \bibnamefont
  {\"{O}zdemir}}, \bibinfo {author} {\bibfnamefont {M.~S.}\ \bibnamefont
  {Kim}}, \ and\ \bibinfo {author} {\bibfnamefont {S.~A.}\ \bibnamefont
  {Maier}},\ }\bibfield  {title} {\emph {\bibinfo {title} {Observation of
  quantum interference in the plasmonic {Hong-Ou-Mandel} effect},}\ }\href@noop
  {} {\bibfield  {journal} {\bibinfo  {journal} {Phys. Rev. Applied}\ }\textbf
  {\bibinfo {volume} {1}},\ \bibinfo {pages} {034004} (\bibinfo {year}
  {2014})}\BibitemShut {NoStop}%
\bibitem [{\citenamefont {Cai}\ \emph {et~al.}(2014)\citenamefont {Cai},
  \citenamefont {Li}, \citenamefont {Ren}, \citenamefont {Zou}, \citenamefont
  {Xiong}, \citenamefont {Lei}, \citenamefont {Liu}, \citenamefont {Guo},\ and\
  \citenamefont {Guo}}]{Cai2014}%
  \BibitemOpen
  \bibfield  {author} {\bibinfo {author} {\bibfnamefont {Y.-J.}\ \bibnamefont
  {Cai}}, \bibinfo {author} {\bibfnamefont {M.}~\bibnamefont {Li}}, \bibinfo
  {author} {\bibfnamefont {X.-F.}\ \bibnamefont {Ren}}, \bibinfo {author}
  {\bibfnamefont {C.-L.}\ \bibnamefont {Zou}}, \bibinfo {author} {\bibfnamefont
  {X.}~\bibnamefont {Xiong}}, \bibinfo {author} {\bibfnamefont {H.-L.}\
  \bibnamefont {Lei}}, \bibinfo {author} {\bibfnamefont {B.-H.}\ \bibnamefont
  {Liu}}, \bibinfo {author} {\bibfnamefont {G.-P.}\ \bibnamefont {Guo}}, \ and\
  \bibinfo {author} {\bibfnamefont {G.-C.}\ \bibnamefont {Guo}},\ }\bibfield
  {title} {\emph {\bibinfo {title} {High-visibility on-chip quantum
  interference of single surface plasmons},}\ }\href@noop {} {\bibfield
  {journal} {\bibinfo  {journal} {Phys. Rev. Applied}\ }\textbf {\bibinfo
  {volume} {2}},\ \bibinfo {pages} {014004} (\bibinfo {year}
  {2014})}\BibitemShut {NoStop}%
\bibitem [{\citenamefont {Fujii}\ \emph {et~al.}(2014)\citenamefont {Fujii},
  \citenamefont {Fukuda},\ and\ \citenamefont {Inoue}}]{Fujii2014}%
  \BibitemOpen
  \bibfield  {author} {\bibinfo {author} {\bibfnamefont {G.}~\bibnamefont
  {Fujii}}, \bibinfo {author} {\bibfnamefont {D.}~\bibnamefont {Fukuda}}, \
  and\ \bibinfo {author} {\bibfnamefont {S.}~\bibnamefont {Inoue}},\ }\bibfield
   {title} {\emph {\bibinfo {title} {Direct observation of bosonic quantum
  interference of surface plasmon polaritons using photon-number-resolving
  detectors},}\ }\href@noop {} {\bibfield  {journal} {\bibinfo  {journal}
  {Phys. Rev. B}\ }\textbf {\bibinfo {volume} {90}},\ \bibinfo {pages} {085430}
  (\bibinfo {year} {2014})}\BibitemShut {NoStop}%
\bibitem [{\citenamefont {Fakonas}\ \emph {et~al.}(2015)\citenamefont
  {Fakonas}, \citenamefont {Mitskovets},\ and\ \citenamefont
  {Atwater}}]{Fakonas2015}%
  \BibitemOpen
  \bibfield  {author} {\bibinfo {author} {\bibfnamefont {J.~S.}\ \bibnamefont
  {Fakonas}}, \bibinfo {author} {\bibfnamefont {A.}~\bibnamefont {Mitskovets}},
  \ and\ \bibinfo {author} {\bibfnamefont {H.~A.}\ \bibnamefont {Atwater}},\
  }\bibfield  {title} {\emph {\bibinfo {title} {Path entanglement of surface
  plasmons},}\ }\href@noop {} {\bibfield  {journal} {\bibinfo  {journal} {New
  J. Phys.}\ }\textbf {\bibinfo {volume} {17}},\ \bibinfo {pages} {023002}
  (\bibinfo {year} {2015})}\BibitemShut {NoStop}%
\bibitem [{\citenamefont {van Loock}(2011)}]{Loock2011}%
  \BibitemOpen
  \bibfield  {author} {\bibinfo {author} {\bibfnamefont {P.}~\bibnamefont {van
  Loock}},\ }\bibfield  {title} {\emph {\bibinfo {title} {Optical hybrid
  approaches to quantum information},}\ }\href {\doibase
  10.1002/lpor.201000005} {\bibfield  {journal} {\bibinfo  {journal} {Laser \&
  Photonics Reviews}\ }\textbf {\bibinfo {volume} {5}},\ \bibinfo {pages} {167}
  (\bibinfo {year} {2011})}\BibitemShut {NoStop}%
\bibitem [{\citenamefont {He}\ \emph {et~al.}(2007)\citenamefont {He},
  \citenamefont {Bergou},\ and\ \citenamefont {Wang}}]{He2007}%
  \BibitemOpen
  \bibfield  {author} {\bibinfo {author} {\bibfnamefont {B.}~\bibnamefont
  {He}}, \bibinfo {author} {\bibfnamefont {J.~A.}\ \bibnamefont {Bergou}}, \
  and\ \bibinfo {author} {\bibfnamefont {Z.}~\bibnamefont {Wang}},\ }\bibfield
  {title} {\emph {\bibinfo {title} {Implementation of quantum operations on
  single-photon qudits},}\ }\href@noop {} {\bibfield  {journal} {\bibinfo
  {journal} {Phys. Rev. A}\ }\textbf {\bibinfo {volume} {76}},\ \bibinfo
  {pages} {042326} (\bibinfo {year} {2007})}\BibitemShut {NoStop}%
\bibitem [{\citenamefont {Leonhardt}(2003)}]{Leonhardt2003}%
  \BibitemOpen
  \bibfield  {author} {\bibinfo {author} {\bibfnamefont {U.}~\bibnamefont
  {Leonhardt}},\ }\bibfield  {title} {\emph {\bibinfo {title} {Quantum physics
  of simple optical instruments},}\ }\href@noop {} {\bibfield  {journal}
  {\bibinfo  {journal} {Rep. Prog. Phys.}\ }\textbf {\bibinfo {volume} {66}},\
  \bibinfo {pages} {1207} (\bibinfo {year} {2003})}\BibitemShut {NoStop}%
\bibitem [{\citenamefont {Williamson}(1937)}]{Williamson1937}%
  \BibitemOpen
  \bibfield  {author} {\bibinfo {author} {\bibfnamefont {J.}~\bibnamefont
  {Williamson}},\ }\bibfield  {title} {\emph {\bibinfo {title} {Quasi-unitary
  matrices},}\ }\href@noop {} {\bibfield  {journal} {\bibinfo  {journal} {Duke
  Math. J}\ }\textbf {\bibinfo {volume} {3}},\ \bibinfo {pages} {715} (\bibinfo
  {year} {1937})}\BibitemShut {NoStop}%
\bibitem [{Note2()}]{Note2}%
  \BibitemOpen
  \bibinfo {note} {We use a slightly different definition for the matrix $H$ in
  this connection: $H=iG \protect \mathrm {ln}S$.}\BibitemShut {Stop}%
\bibitem [{\citenamefont {Andersen}\ \emph {et~al.}(2016)\citenamefont
  {Andersen}, \citenamefont {Gehring}, \citenamefont {Marquardt},\ and\
  \citenamefont {Leuchs}}]{Andersen2016}%
  \BibitemOpen
  \bibfield  {author} {\bibinfo {author} {\bibfnamefont {U.~L.}\ \bibnamefont
  {Andersen}}, \bibinfo {author} {\bibfnamefont {T.}~\bibnamefont {Gehring}},
  \bibinfo {author} {\bibfnamefont {C.}~\bibnamefont {Marquardt}}, \ and\
  \bibinfo {author} {\bibfnamefont {G.}~\bibnamefont {Leuchs}},\ }\bibfield
  {title} {\emph {\bibinfo {title} {30 years of squeezed light generation},}\
  }\href@noop {} {\bibfield  {journal} {\bibinfo  {journal} {Phys. Scr.}\
  }\textbf {\bibinfo {volume} {91}},\ \bibinfo {pages} {053001} (\bibinfo
  {year} {2016})}\BibitemShut {NoStop}%
\bibitem [{\citenamefont {Peres}(1990)}]{Peres1990}%
  \BibitemOpen
  \bibfield  {author} {\bibinfo {author} {\bibfnamefont {A.}~\bibnamefont
  {Peres}},\ }\bibfield  {title} {\emph {\bibinfo {title} {Neumark's theorem
  and quantum inseparability},}\ }\href@noop {} {\bibfield  {journal} {\bibinfo
   {journal} {Found. Phys.}\ }\textbf {\bibinfo {volume} {20}},\ \bibinfo
  {pages} {1441} (\bibinfo {year} {1990})}\BibitemShut {NoStop}%
\bibitem [{Note3()}]{Note3}%
  \BibitemOpen
  \bibinfo {note} {By $\left (a:b,c:d\right )$ we denote the submatrix
  consisting of the intersection of rows $a$ to $b$ and columns $c$ to $d$ of
  the original matrix}\BibitemShut {NoStop}%
\bibitem [{\citenamefont {Nielsen}\ and\ \citenamefont
  {Chuang}(2000)}]{Nielsen}%
  \BibitemOpen
  \bibfield  {author} {\bibinfo {author} {\bibfnamefont {M.}~\bibnamefont
  {Nielsen}}\ and\ \bibinfo {author} {\bibfnamefont {I.}~\bibnamefont
  {Chuang}},\ }\href@noop {} {\emph {\bibinfo {title} {Quantum Computation and
  Quantum Information}}}\ (\bibinfo  {publisher} {Cambridge University Press},\
  \bibinfo {year} {2000})\BibitemShut {NoStop}%
\bibitem [{\citenamefont {He}\ and\ \citenamefont {Bergou}(2006)}]{He2006}%
  \BibitemOpen
  \bibfield  {author} {\bibinfo {author} {\bibfnamefont {B.}~\bibnamefont
  {He}}\ and\ \bibinfo {author} {\bibfnamefont {J.~A.}\ \bibnamefont
  {Bergou}},\ }\bibfield  {title} {\emph {\bibinfo {title} {A general approach
  to physical realization of unambiguous quantum-state discrimination},}\
  }\href {\doibase https://doi.org/10.1016/j.physleta.2006.03.076} {\bibfield
  {journal} {\bibinfo  {journal} {Phys. Lett. A}\ }\textbf {\bibinfo {volume}
  {356}},\ \bibinfo {pages} {306 } (\bibinfo {year} {2006})}\BibitemShut
  {NoStop}%
\bibitem [{\citenamefont {Tabia}(2012)}]{Tabia2012}%
  \BibitemOpen
  \bibfield  {author} {\bibinfo {author} {\bibfnamefont {G.~N.~M.}\
  \bibnamefont {Tabia}},\ }\bibfield  {title} {\emph {\bibinfo {title}
  {Experimental scheme for qubit and qutrit symmetric informationally complete
  positive operator-valued measurements using multiport devices},}\ }\href@noop
  {} {\bibfield  {journal} {\bibinfo  {journal} {Phys. Rev. A}\ }\textbf
  {\bibinfo {volume} {86}},\ \bibinfo {pages} {062107} (\bibinfo {year}
  {2012})}\BibitemShut {NoStop}%
\bibitem [{\citenamefont {Bian}\ \emph {et~al.}(2015)\citenamefont {Bian},
  \citenamefont {Li}, \citenamefont {Qin}, \citenamefont {Zhan}, \citenamefont
  {Zhang}, \citenamefont {Sanders},\ and\ \citenamefont {Xue}}]{Bian2015}%
  \BibitemOpen
  \bibfield  {author} {\bibinfo {author} {\bibfnamefont {Z.}~\bibnamefont
  {Bian}}, \bibinfo {author} {\bibfnamefont {J.}~\bibnamefont {Li}}, \bibinfo
  {author} {\bibfnamefont {H.}~\bibnamefont {Qin}}, \bibinfo {author}
  {\bibfnamefont {X.}~\bibnamefont {Zhan}}, \bibinfo {author} {\bibfnamefont
  {R.}~\bibnamefont {Zhang}}, \bibinfo {author} {\bibfnamefont {B.~C.}\
  \bibnamefont {Sanders}}, \ and\ \bibinfo {author} {\bibfnamefont
  {P.}~\bibnamefont {Xue}},\ }\bibfield  {title} {\emph {\bibinfo {title}
  {Realization of single-qubit positive-operator-valued measurement via a
  one-dimensional photonic quantum walk},}\ }\href {\doibase
  10.1103/PhysRevLett.114.203602} {\bibfield  {journal} {\bibinfo  {journal}
  {Phys. Rev. Lett.}\ }\textbf {\bibinfo {volume} {114}},\ \bibinfo {pages}
  {203602} (\bibinfo {year} {2015})}\BibitemShut {NoStop}%
\bibitem [{\citenamefont {Bent}\ \emph {et~al.}(2015)\citenamefont {Bent},
  \citenamefont {Qassim}, \citenamefont {Tahir}, \citenamefont {Sych},
  \citenamefont {Leuchs}, \citenamefont {S{\'a}nchez-Soto}, \citenamefont
  {Karimi},\ and\ \citenamefont {Boyd}}]{Bent2015}%
  \BibitemOpen
  \bibfield  {author} {\bibinfo {author} {\bibfnamefont {N.}~\bibnamefont
  {Bent}}, \bibinfo {author} {\bibfnamefont {H.}~\bibnamefont {Qassim}},
  \bibinfo {author} {\bibfnamefont {A.~A.}\ \bibnamefont {Tahir}}, \bibinfo
  {author} {\bibfnamefont {D.}~\bibnamefont {Sych}}, \bibinfo {author}
  {\bibfnamefont {G.}~\bibnamefont {Leuchs}}, \bibinfo {author} {\bibfnamefont
  {L.~L.}\ \bibnamefont {S{\'a}nchez-Soto}}, \bibinfo {author} {\bibfnamefont
  {E.}~\bibnamefont {Karimi}}, \ and\ \bibinfo {author} {\bibfnamefont {R.~W.}\
  \bibnamefont {Boyd}},\ }\bibfield  {title} {\emph {\bibinfo {title}
  {Experimental realization of quantum tomography of photonic qudits via
  symmetric informationally complete positive operator-valued measures},}\
  }\href@noop {} {\bibfield  {journal} {\bibinfo  {journal} {Phys. Rev. X}\
  }\textbf {\bibinfo {volume} {5}},\ \bibinfo {pages} {041006} (\bibinfo {year}
  {2015})}\BibitemShut {NoStop}%
\bibitem [{\citenamefont {Sosa-Martinez}\ \emph {et~al.}(2017)\citenamefont
  {Sosa-Martinez}, \citenamefont {Lysne}, \citenamefont {Baldwin},
  \citenamefont {Kalev}, \citenamefont {Deutsch},\ and\ \citenamefont
  {Jessen}}]{Sosa2017}%
  \BibitemOpen
  \bibfield  {author} {\bibinfo {author} {\bibfnamefont {H.}~\bibnamefont
  {Sosa-Martinez}}, \bibinfo {author} {\bibfnamefont {N.~K.}\ \bibnamefont
  {Lysne}}, \bibinfo {author} {\bibfnamefont {C.~H.}\ \bibnamefont {Baldwin}},
  \bibinfo {author} {\bibfnamefont {A.}~\bibnamefont {Kalev}}, \bibinfo
  {author} {\bibfnamefont {I.~H.}\ \bibnamefont {Deutsch}}, \ and\ \bibinfo
  {author} {\bibfnamefont {P.~S.}\ \bibnamefont {Jessen}},\ }\bibfield  {title}
  {\emph {\bibinfo {title} {Experimental study of optimal measurements for
  quantum state tomography},}\ }\href@noop {} {\bibfield  {journal} {\bibinfo
  {journal} {Phys. Rev. Lett.}\ }\textbf {\bibinfo {volume} {119}},\ \bibinfo
  {pages} {150401} (\bibinfo {year} {2017})}\BibitemShut {NoStop}%
\bibitem [{\citenamefont {Hofmann}\ and\ \citenamefont
  {Takeuchi}(2002)}]{Hofmann2002}%
  \BibitemOpen
  \bibfield  {author} {\bibinfo {author} {\bibfnamefont {H.~F.}\ \bibnamefont
  {Hofmann}}\ and\ \bibinfo {author} {\bibfnamefont {S.}~\bibnamefont
  {Takeuchi}},\ }\bibfield  {title} {\emph {\bibinfo {title} {Quantum phase
  gate for photonic qubits using only beam splitters and postselection},}\
  }\href@noop {} {\bibfield  {journal} {\bibinfo  {journal} {Phys. Rev. A}\
  }\textbf {\bibinfo {volume} {66}},\ \bibinfo {pages} {024308} (\bibinfo
  {year} {2002})}\BibitemShut {NoStop}%
\end{thebibliography}%

\end{document}